\documentclass[a4paper,11pt]{article}
\input{paper.sty}

\title{Bayesian sensitivity of binary pulsars to ultra-light dark matter}

\author[a,b]{Pavel K\r{u}s,}
\author[a,c,d]{Diana L\'opez Nacir}
\author[a]{and Federico R.~Urban}
\affiliation[a]{CEICO, FZU--Institute of Physics of the Czech Academy of Sciences\\Na Slovance 2, 182 00 Praha 8, Czech Republic}
\affiliation[b]{Charles University, Faculty of Mathematics and Physics, Institute of Theoretical Physics \\ V Hole\v{s}ovi\v{c}k\'ach 2, 180 00 Praha 8, Czech Republic}
\affiliation[c]{Universidad de Buenos Aires, Facultad de Ciencias Exactas y Naturales, Departamento de Física. Buenos Aires, Argentina.}
\affiliation[d]{CONICET - Universidad de Buenos Aires, Instituto de Física de Buenos Aires (IFIBA). Buenos Aires, Argentina}

\emailAdd{pavel.kus@fzu.cz}
\emailAdd{dnacir@df.uba.ar}
\emailAdd{federico.urban@fzu.cz}

\abstract{Ultra-light dark matter perturbs the orbital motion of binary pulsars, in particular by causing peculiar time variations of a binary's orbital parameters, which then induce variations in the pulses' times-of-arrival. Binary pulsars have therefore been shown to be promising detectors of ultra-light dark matter. To date, the sensitivity of binary pulsars to ultra-light dark matter has only been studied for dark matter masses in a narrow resonance band around a multiple of the binary pulsar orbital frequency. In this study we devise a two-step, bayesian method that enables us to compute semi-analytically the sensitivity for all masses, also away from the resonance, and to combine several observed binaries into one global sensitivity curve. We then apply our method to the case of a universal, linearly-coupled, scalar ultra-light dark matter. We find that with next-generation radio observatories the sensitivity to the ultra-light dark matter coupling will surpass that of solar-system constraints for a decade in mass around \(m\sim10^{-21}\,\text{eV}\), even beyond resonance.}

\begin{document}
\maketitle
\flushbottom

%===============================================================================
% BODY
%===============================================================================

%-------------------------------------------------------------------------------
\section{Introduction}
\label{sec:intro}
%-------------------------------------------------------------------------------

Binary pulsars, namely astrophysical binary systems in which at least one of the two orbiting bodies is a pulsar, are outstandingly precise clocks, and as such they have been used in a wide variety of applications---see \cite{Lorimer:2008se} for a comprehensive review. Among those applications, binary pulsars can act as potential detectors for dark matter whose mass lies in the lower end of the ultra-light dark matter (ULDM) range: \(10^{-23}\,\text{eV}\lesssim m \lesssim 10^{-18}\,\text{eV}\). In the late Universe, ULDM with such a mass behaves like a coherent superposition of oscillators all with the same inverse frequency of about $1.15\,\mathrm{hours}\times(10^{-18}\,\mathrm{eV}/m)$, which roughly corresponds to the span of periods of observed binary pulsars. Two interesting phenomena take place when the orbital period of the system, or any of its harmonics, is in resonance with the ULDM oscillations. First, because the ULDM oscillations perturb the gravitational potential in which the binary system is immersed, they cause secular drifts in the orbital motion, which however are unlikely to be observed. Second, if the ULDM carries a (fifth) force that couples directly to the bodies of the binary system, it will bring about further secular drifts in the orbital parameters that can be potentially detected.\footnote{The times of arrival of pulses of pulsars, whether isolated or in binary systems, are also directly sensitive to ULDM oscillations, a fact that can be exploited by pulsar-timing arrays~\cite{Khmelnitsky:2013lxt,Armaleo:2020yml}.} Such resonant effects have been studied for ULDM of spin 0, spin 1 and spin 2~\cite{Blas:2016ddr,Blas:2019hxz,LopezNacir:2018epg,Armaleo:2019gil}, and in all cases binary pulsars have been shown to be promising detectors for such kinds of ULDM.

These results however are limiting in three respects. First, resonances appear only for a very narrow range of ULDM mass \(m\) peaked at a given harmonic of the orbital period; for the linear direct coupling we consider here \(m\approx k\omega_b\) where \(k\in \mathbb{N}\) is the harmonic number and \(\omega_b=2\pi/P_b\) is the orbital frequency and \(P_b\) the orbital period. This means that each binary system can probe a very small parameter space in ULDM mass, or, in other words, that only very few, if any, binary pulsars will be sensitive to ULDM. Second, for the most-studied case of a universally-coupled spin 0, the resonant secular drift on the period \(P_b\)---which is often the best-measured parameter---vanishes for systems with small eccentricities \(e\rightarrow0\), which are the majority (e.g.\ more than 80\% in the ATNF catalogue~\cite{Manchester:2004bp} have \(e<0.001\)).\footnote{The secular drifts for non-universally-coupled scalars does not vanish with \(e\rightarrow0\), but are further suppressed by the presumably small asymmetry in couplings between the two orbiting bodies.} Hence, in this case we are limited to an even smaller sub-sample of all observed binary pulsars. Third, so far the sensitivity of binary pulsars to ULDM has only been estimated by requiring that the ULDM effects be within the errors of the measured orbital parameters, thus ignoring the time behaviour of the ULDM perturbations on the orbital parameters.

In this work we develop a new method to estimate the sensitivity of binary pulsars as detectors of ULDM. For practical purposes here we limit ourselves to the universally-coupled linear scalar case, but our results are readily generalised to other spins and types of coupling. Our proposal draws from existing methods used to determine the sensitivity of pulsar-timing arrays to continuous gravitational waves, and takes advantage of the full time evolution of the ULDM perturbations to the binary system. In this way we are able to improve on existing results in three different ways. First, for a given system, we are able to compute semi-analytically the sensitivity to ULDM for all values of the ULDM mass, that is, also away from the resonance. Second, since we are not limited to resonances, we will see that the effects of ULDM do not vanish for \(e\rightarrow0\), only the resonances do; hence, even for a universally-coupled scalar ULDM, all binary pulsars can act as ULDM detectors. Third, with our method we can straightforwardly combine multiple binary pulsars, each of which spans the entire mass range of interest. This vastly improves the sensitivity to ULDM and can tighten the constraint on the ULDM parameter space should no signal be found.

The rest of this paper is organised as follows: in \autoref{sec:pert} we introduce the post-Keplerian, osculating-orbits formalism and derive the effects of ULDM on the six orbital parameters. In \autoref{sec:est} we describe our method to estimate the sensitivity. We present our results in \autoref{sec:res} and conclude with an outlook for future work in \autoref{sec:end}. In the Appendix we collect several useful formulas and additional tests and cross-checks of our method. Throughout the paper we use units for which \(c=1\) and $\hbar = 1$.

%-------------------------------------------------------------------------------
\section{The perturbed orbital motion}
\label{sec:pert}
%-------------------------------------------------------------------------------

%-------------------------------------------------------------------------------
\subsection{Timing models}
\label{ssec:tim}

The phase \(N\) of the pulsar at the moment at which the pulse is emitted is related to the proper time as measured by a hypothetical clock on the pulsar \(T\) by
\begin{equation}\label{Npulse}
    N = N_0+ \nu T + \frac12\frac{\mathrm{d}\nu}{\mathrm{d}T}T^2 + \frac16\frac{\mathrm{d}^2\nu}{\mathrm{d}T^2}T^3 \,,
\end{equation}
where $\nu$ is the spin frequency of the pulsar and $N_0$ an arbitrary constant, see~\cite{Lorimer:2004}. The pulsar's proper time $T$ is related to the time of arrival of each pulse at the detector on Earth through a timing model. In this work we focus on two timing models, the BT model of~\cite{Teukolsky1976} and the ELL1 model of~\cite{Lange:2001rn}, which better describes near-circular orbits.

\begin{figure}
    \centering
    \includegraphics[width=0.8\textwidth]{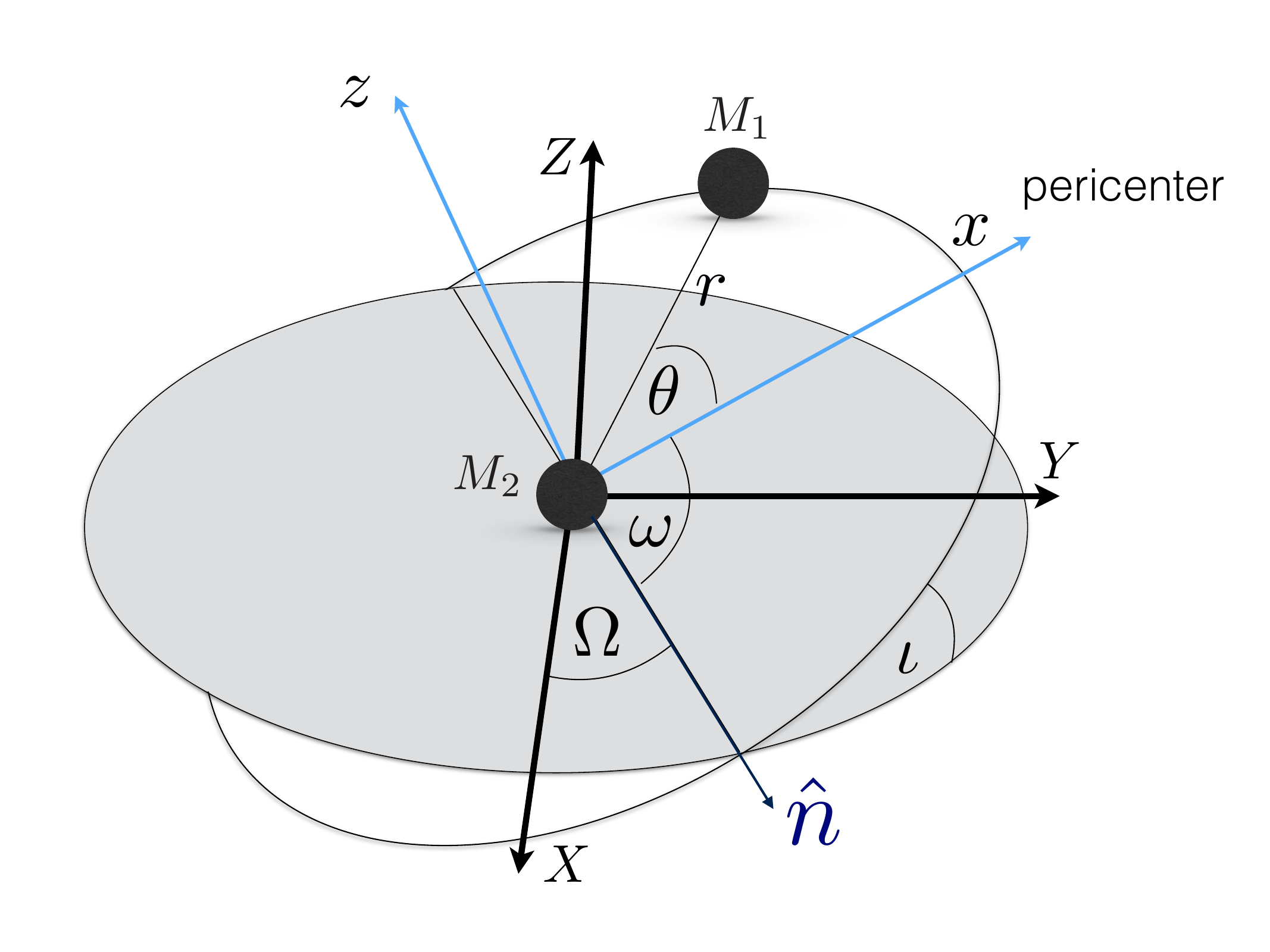}
    \caption{Description of Keplerian orbits in terms of the orbital elements viewed in the fundamental reference frame \((X, Y, Z)\). The Cartesian orbital frame \((x, y, z)\) and the polar one \((r, \theta, z)\) are also shown (centred on \(M_2\) for convenience).}
    \label{fig:orbit}
\end{figure}

For non-relativistic orbits the relation between proper and Earth's times can be written as
\begin{equation}\label{DelayTeu}
    t=T+\alpha_b (\cos E-e)+\beta_b\sin E \,,
\end{equation}
where we defined
\begin{subequations}\label{defalphabetab}
\begin{align}
    \alpha_b &\deq x\sin\omega \,,~~\beta_b \deq (1-e^2)^{1/2} x \cos\omega \,, \\
    x &\deq a_1 s \,,~~s\deq\sin\iota \,,~~a_1 \deq \frac{M_2}{M_1+M_2}a \,,
\end{align}
\end{subequations}
where \(a\) is the semi-major axis of the binary, \(e\) its eccentricity, \(\iota\) its inclination, \(\omega\) the argument of periapsis (not to be confused with the orbital frequency \(\omega_b\)), and \(M_1\) and \(M_2\) are the masses of the pulsars and its companion, respectively---the Euler angles \(\iota\) and \(\omega\), together with the longitude of the ascending node \(\Omega\) define the rotation of the \((x,y,z)\) system with respect to the observer's system \((X,Y,Z)\) (in which the observer is at \(Z=-\infty\)), see \autoref{fig:orbit}.\footnote{Strictly speaking, $t$ is not the time measured by an observer on Earth, but it corresponds to the so-called ``infinite-frequency barycentre arrival time’’ which is a hypothetical time of arrival at the barycentre of the solar system that includes corrections due to the motion of the Earth with respect to the barycentre, the gravitational redshift in the solar system and the effect of interstellar dispersion (see~\cite{Teukolsky1976} for an introduction). Here we are assuming that such effects can be computed so that $t$ is a known function of the arrival time at Earth.} The eccentric anomaly $E$ satisfies Kepler's equation
\begin{equation}\label{eq:defTh}
    E-e\sin E = \int_{T_0}^T \omega_b\, \di T' \deq \Theta \,,
\end{equation}
where \(T_0\) is the proper time of periapsis passage.\footnote{Notice that here we use a different notation than in~\cite{Teukolsky1976}, namely $\Theta=\sigma + \int_0^T \omega_b \di T'$.} In order to make contact with the literature, we then define
\begin{align}\label{eq:etaDef}
    \eta_b \deq \beta_b+\gamma_b \,,
\end{align}
where \(\gamma_b\) is the Einstein delay; in practice this term is presently indistinguishable from $\beta_b$, which means that \(\eta_b\approx\beta_b\). Lastly, in this work we always ignore the Shapiro delay term in the timing model.

With the timing model at hand we can perturbatively solve for $N$ as a function of $t$ and the timing-model parameters to obtain
\begin{equation}\label{Npulset}
    N=N_0+\nu t+\frac{\dot\nu}{2}t^2+ \frac{\ddot\nu}{6}t^3 - \nu f (E') - \dot\nu t f (E') + \frac{\nu\omega_b}{1-e\cos E'} f(z)\frac{df(z)}{dz}\Big{ |}_{z=E'} \,,
\end{equation}
where we defined the function
\begin{equation}
    f(z)\deq\alpha_b (\cos z-e)+\eta_b\sin z \,,
\end{equation}
and a new eccentric anomaly \(E'\) as
\begin{equation}\label{eq:defThp}
    E'-e\sin E'=\int_{T_0}^t \omega_b \,\di t' \deq\Theta' \,.
\end{equation}
\autoref{Npulset} defines the BT timing model, for which we can equivalently use the parameter sets \(\{a,e,\omega,s,T_0\}\) or \(\{a,e,\alpha_b,\eta_b,\Theta'\}\). On top of these, there are the parameters \(N_0\), \(\nu\), \(\dot\nu\) and \(\ddot\nu\) that are related to the pulsar's own rotation (and are common to all timing models).

If the orbit is nearly circular, $e\rightarrow0$, $T_0$ and $\omega$ are ill-defined because there is no periapsis. In this case we can define a different set of variables, the Laplace-Lagrange parameters \(\eta\) (not to be confused with parameter \(\eta_b\) defined in \autoref{eq:etaDef}) and \(\kappa\), as
\begin{subequations}
\begin{align}
    \eta &\deq e\sin\omega \,,\label{eq:defEta}\\
    \kappa &\deq e\cos\omega \,,\label{eq:defKappa}
\end{align}
\end{subequations}
and we can replace $T_0$ with $T_\text{asc}$, which stands for the time of ascending node. In analogy with the definition \autoref{eq:defTh} we write
\begin{align}\label{eq:psiDef}
    \Psi &\deq \int_{T_\text{asc}}^{T} \omega_b \, \di T' = E-e\sin E + \omega \,,
\end{align}
with \(\omega = \int_{T_\text{asc}}^{T_0} \omega_b \, \di T'\).\footnote{Notice that here we call $\Psi$ to denote what is often called $\Phi$ in the literature (see for instance~\cite{Lange:2001rn,Hobbs:2006cd}), because in our notation $\Phi$ stands for the ULDM field.} In order to make contact with the new eccentric anomaly \(E'\) we can finally define a new variable $\Psi'$ through
\begin{align}
    \Psi' \deq \int_{T_\text{asc}}^t \omega_b \, \di t' = E'-e\sin E' + \omega \,.
\end{align}
With these parameters, the relationship \autoref{Npulset} becomes
\begin{equation}
\label{NpulsetELL}
    N=N_0+\nu t+\frac{\dot\nu}{2}t^2+ \frac{\ddot\nu}{6}t^3-\dot\nu x g(\Psi')t-\nu\,g(\Psi')+\nu\omega_b\,g(z) \frac{dg(z)  }{dz}\Big{ |}_{z=\Psi'} \,,
\end{equation}
where now
\begin{equation}
    g(z) \deq -x\frac{ \eta } {2}\cos 2 z -x\frac{3 \eta }{2}+x\frac{\kappa} {2} \sin 2 z +x \sin z \,.
\end{equation}
\autoref{NpulsetELL} defines the ELL1 timing model~\cite{Lange:2001rn}, that is, the set \(\{x,\eta,\kappa,s,T_\text{asc}\}\) or with \(T_\text{asc}\rightarrow\Psi'\).
%-------------------------------------------------------------------------------
\subsection{Ultra-light dark matter as a perturbing force}
\label{ssec:uldm}

The unperturbed orbital motion of two bodies in the binary is completely characterised by six orbital parameters: $a$, $e$, $\iota$, $\omega$, $T_0$ and the longitude of the ascending node $\Omega$. The ULDM field \(\Phi(t)\), owing to its interaction with the binary system, perturbs the system's orbital motion and modifies the timing models of \autoref{Npulset} or \autoref{NpulsetELL} that describe it. To illustrate our method, here we consider an effective, universal, direct linear coupling between the ULDM field and the two bodies in the binary system. The coupling is characterised by a single parameter $\alpha$ for which we can perturbatively write
\begin{equation}\label{massdep}
    M_A^{(\alpha)}(\Phi) = M_A(1+\alpha\Phi) \,,
\end{equation}
where $A\in\{1,2\}$, with $M_1$ being the pulsar's mass and $M_2$ the mass of the companion.\footnote{If the scalar theory is endowed with a, e.g.\ \(Z_2\) symmetry, the first interaction term would be quadratic in the field \(\Phi\). We will comment on this case in \autoref{sec:end}. Notice that we also do not take into account non-perturbative effects such as scalarisation.}

The ULDM field can be considered homogeneous within its de Broglie wavelength (or coherence length), as determined by
\begin{equation}\label{deBroglie}
    \lambda_{\mathrm{dB}} \sim 1.3\times 10^{12}~\mathrm{km} \left( \frac{10^{-3}}{v_0} \right) \left( \frac{10^{-18}~\mathrm{eV} }{m} \right) \,,
\end{equation}
where \(v_0\) is the typical ULDM velocity in the Milky Way halo. However, its amplitude and phase are expected to show large oscillations from binary system to binary system. Therefore, we describe the ULDM field as~\cite{Foster:2017hbq}
\begin{equation}\label{Phi}
    \Phi=\Phi_0 \varrho \cos(m t+\Upsilon) \,,
\end{equation}
where, assuming that each binary system is in a different coherence patch, $\Upsilon \in [0,2\pi)$ is a random phase,
\begin{equation}\label{Phi0}
    \Phi_0 \deq \frac{\sqrt{2\rho_\text{DM}}}{m} \,,
\end{equation}
the local ULDM density is taken to be $\rho_\text{DM} = 0.3$ $\mathrm{GeV/cm^3}$ and $\varrho\geq0$ is a random variable drawn from the Rayleigh distribution 
\begin{equation}\label{Pr}
    P(\varrho) = 2 \varrho e^{-\varrho^2} \,.
\end{equation}
The stochastic character of the ULDM field comes about because pulsar-timing observation campaigns typically last up to a few decades, which is a much shorter time than the ULDM coherence time
\begin{equation}\label{tcoh}
    t_{\mathrm{coh}} \sim 65~\mathrm{y} \left( \frac{10^{-3}}{v_0} \right)^2 \left(\frac{10^{-18}~\mathrm{eV} } {m} \right) \,;
\end{equation}
indeed, only by sampling the field during several \(t_\mathrm{coh}\) does the amplitude converge to \(\varrho\rightarrow1\).

The perturbed, non-relativistic Keplerian equations of motion of a binary system interacting with ULDM are given by
\begin{subequations}
\begin{align}
    \ddot{\vec{R}} =&-\alpha\dot\Phi\dot{\vec{R}} \,, \label{eq:Rcm}\\
    \ddot{\vec{r}} =&-(1+\alpha \Phi)\frac{G M_T\vec{r}}{r^3} -\alpha \dot{\Phi}\,\dot{\vec{r}} = -\frac{G M_T\vec{r}}{r^3}+\vec{F} \,, \label{eq:r12}
\end{align}
\end{subequations}
where a dot denotes a derivative with respect to (coordinate) time, $M_T \deq M_1+M_2$, ${\vec{R}}$ is the binary barycentre position defined using the unperturbed masses, ${\vec{r}}$ is the relative coordinate with origin in the companion body with mass $M_2$ (see~\cite{Blas:2019hxz}) that defines the polar coordinate system \((\hat r,\hat \theta,\hat z\)) of the orbital motion. In the last equality we defined $\vec{F}$ which we will refer to as the ``perturbing force''.

%-------------------------------------------------------------------------------
\subsection{Osculating orbits}
\label{ssec:orb}

The perturbing force $\vec F$ defined in \autoref{eq:r12} leads to deviations from the Kepler motion that can be studied using the osculating orbits formalism as described in~\cite{Danby:1970}. After some algebra (see \autoref{app:sec}), in the BT model we obtain
\begin{subequations}\label{OSCorbits}
\begin{align}
    \frac{\dot{a}}{a} =& -\frac{2\alpha \Phi e\omega_b}{\sqrt{1-e^2}} \frac{a^2}{r^2}\sin\theta - \frac{2\alpha \dot{\Phi}(1+e^2+2e\cos\theta)}{1-e^2} \,, \label{eqa}\\
    \dot{e} =& -\alpha \Phi\sqrt{1-e^2}\omega_b\frac{a^2}{r^2}\sin\theta -2\alpha \dot{\Phi}(\cos\theta+e) \,, \label{eqe} \\
    \dot{\omega} =&\, \frac{\alpha \Phi\omega_b\sqrt{1-e^2}}{e} \frac{a^2}{r^2}\cos\theta -\frac{2\alpha \dot{\Phi}}{e} \sin\theta \,, \\
    \dot{\epsilon_1} =&\, \frac{2\alpha\Phi\omega_b}{1-e^2}(1+e\cos\theta) +\frac{2\alpha \dot{\Phi}e}{\sqrt{1-e^2}}\frac{r}{a}\sin\theta +[1-\sqrt{1-e^2}]\dot{\omega} \,, \\
    \dot{\Omega} =&\, \dot{s} = 0 \,,
\end{align}
\end{subequations}
where the parameter $\epsilon_1$ is given by
\begin{equation}\label{eps1}
    \epsilon_1\deq\omega_b(t-T_0)+\omega+\Omega-\int_{T_0}^t \di t' \,\omega_b\,,
\end{equation}
and is related to $\Theta'=\Psi'-\omega$ by $\dot{\epsilon_1}=\dot{\Omega}+\dot{\omega}+\dot{\Theta}'-\omega_b$.\footnote{The parameter $\epsilon_1$ is defined in~\cite{Danby:1970} in order to avoid a term in the equation where $t$ appears as a coefficient. Such a term could produce large quantities when evolving the equations for a long time.} Notice that scalar ULDM does not have any effect on the parameters \(\Omega\) and \(s\), which we can henceforth discard. Integrating \(\dot{\Theta}'\), using \(\omega_b^2=GM_T/a^3\) and allowing for a possible error  and secular evolution of the semi-major axis $a\to a+\dot a (t-T_\text{asc})$, we find
\begin{eqnarray}\label{eq:varThp}
    \delta\Theta' &=&\delta\Theta'_0+\int_{T_0}^{t}\left[\sqrt{\frac{G M_T(1+\alpha\Phi)}{(a+\delta a)^3}}  -\sqrt{\frac{G M_T }{a ^3}}\right]\di t'+2\alpha \omega_b \int_{T_0}^t \Phi\, \di t'\nonumber\\
    &&+\alpha\sum_{n=1}^{\infty} \int_{T_0}^t  \left[A_n \Phi \cos(n \omega_b \tilde t')+ B_n \dot\Phi \sin(n \omega_b \tilde t')\right]\, \di t' \,,
\end{eqnarray}
where $\delta\Theta'_0$ is the error in the determination of $\Theta'$ at $T_0$, $\tilde{t}' \deq t'-T_0$, with $T_0=T_\text{asc}+\omega/\omega_b$, and
\begin{subequations}\label{AnBn}
\begin{align}
    A_n& \deq 4 \omega_b J_n(n e)-2 n \omega_b \frac{1-e^2}{e} J_n'(n e) \,,\\
    B_n& \deq \frac{4}{n}  J_n(n e)+4  \frac{1-e^2}{e} J_n'(n e) \,,
\end{align}
\end{subequations}
where \(J_n(z)\) is a Bessel function of order \(n\) and \(J_n'(z)\) its derivative with respect to its argument; in obtaining these expressions we have made use of the Fourier decomposition of the orbit, see \autoref{app:sec}.

For orbits with small eccentricity that are described by the ELL1 model, at leading order in \(e\rightarrow0\) the osculating orbit equations become
\begin{subequations}
\begin{align}
    \frac{\dot{x}}{x} &= -2\alpha \dot{\Phi} \,, \label{eqaecc}\\
    \frac{\dot{\eta}}{\eta} = \frac{\dot{\kappa}}{\kappa} &= -\frac{\alpha}{e} \left[\Phi\omega_b\sin\omega_b \tilde{t} + 2\dot{\Phi}\cos\omega_b\tilde{t}\right] \,, \label{eqetaecc}\\
    \dot{\Psi'}-\omega_b &= 2\alpha\Phi\omega_b \,.
\end{align}
\end{subequations}
Proceeding as with $\delta\Theta'$, we find
\begin{align}\label{eq:oscPsi}
    \delta\Psi' &= \delta\Psi_\text{asc}' +\int_{T_\text{asc}}^{t}\left[\sqrt{\frac{G M_T(1+\alpha\Phi)}{(a+\delta a)^3}} -\sqrt{\frac{G M_T }{a ^3}}\right] \di t' + 2\alpha\omega_b\int_{T_\text{asc}}^{t} \Phi\,\di t' \,.
\end{align}
 With $\Phi$ from \autoref{Phi}, upon integration we find the expressions for the variations (denoted by $\delta$) of the orbital parameters, which we collect in \autoref{app:bayes}.

%-------------------------------------------------------------------------------
\section{A two-step approach to estimate the sensitivity}
\label{sec:est}
%-------------------------------------------------------------------------------

\paragraph{Step 1: variances} In order to estimate the sensitivity to ULDM we follow the procedure outlined in~\cite{Teukolsky1976}, which we adapt to our needs by incorporating the Bayesian analysis of~\cite{Moore:2014eua}. From \autoref{Npulset} we see that the time-dependent function $N$ is parametrised by $N_0$, $\nu$, $\dot\nu$, $\ddot\nu$ and the orbital parameters, namely $N=N(t,N_0,\nu, ...)$. The first step is to assume that, for each binary system, we have a reasonable ``first guess'' as to what the values of the orbital parameters are, which we denote $\{N_0^{(1)}$, $\nu^{(1)}$, $\dots\}$. This can be for instance obtained by fitting time-of-arrival data with the help of tools such as TEMPO2~\cite{Hobbs:2006cd} or PINT~\cite{Luo:2020ksx}. We can then define the time residuals $R(t)$ as
\begin{equation}\label{eq:residuals}
    -\nu^{(1)} R(t) \deq N(t,N_0,\nu,\ldots)-N(t,N_0^{(1)},\nu^{(1)},\ldots) \,,
\end{equation}
where $N(t,N_0,\nu,\ldots)$ corresponds to the observed time series (or a synthetic one with model parameters \(\{N_0,\nu,\ldots\}\)). Explicitly, the function $R(t)$ reads
\begin{subequations}    
\begin{align}\label{ResBT}
    R(t) &= \delta K-\frac{\partial N}{\partial a}\Big{|}_{1}\frac{\delta a}{\nu}- \frac{\partial N}{\partial e}\Big{|}_{1}\frac{\delta e}{\nu} -\frac{\partial N}{\partial \alpha_b}\Big{|}_{1}\frac{\delta\alpha_b}{\nu}-\frac{\partial N}{\partial \eta_b}\Big{|}_{1}\frac{\delta \eta_b}{\nu}-\frac{\partial N}{\partial \Theta'}\Big{ |}_{1}\frac{\delta\Theta'}{\nu}\,,\quad\mbox{(BT)} \\\label{ResELL1}
    R(t) &= \delta K-\frac{\partial N}{\partial x}\Big{|}_{1}\frac{\delta x}{\nu}-\frac{\partial N}{\partial \eta}\Big{|}_{1}\frac{\delta\eta}{\nu}-\frac{\partial N}{\partial \kappa}\Big{|}_{1}\frac{\delta\kappa}{\nu} -\frac{\partial N}{\partial s}\Big{|}_{1}\frac{\delta s}{\nu}-\frac{\partial N}{\partial \Psi'}\Big{ |}_{1}\frac{\delta\Psi'}{\nu}\,,\quad\mbox{(ELL1)}
\end{align}
\end{subequations}
where
\begin{equation}\label{eq:deK}
    \delta K \deq -\frac{\delta N_0}{\nu}-\frac{\delta\nu }{\nu}t -\frac{\delta\dot\nu}{2\nu}t^2- \frac{\delta\ddot\nu}{6 \nu}t^3 \,.
\end{equation}
In order to simplify the notation below we define a vector \(\mathsf{\delta S}\) containing the model parameters of the residuals that can in principle be measured for both the BT and ELL1 timing models, namely \(\mathsf{\delta S}_\text{BT}\deq \{\delta K\,,\delta a\,,\delta e\,,\delta \alpha_b\,,\delta \eta_b\,,\delta\Theta'\}\) and \(\mathsf{\delta S}_\text{ELL1}\deq \{\delta K\,,\delta s\,,\delta x\,,\delta \eta\,,\delta \kappa\,,\delta\Psi'\}\), respectively. In what follows we will neglect the contribution of the term with $\dot\nu$ everywhere in \autoref{ResBT} except in $\delta K$.

We now assume that the observations are made at regular time intervals at a rate $n_c$ for a duration $d$ such that $P_b\ll d\ll T_\text{obs}$ where $T_\text{obs}$ is the total time of observation. We also assume that each observation has a constant variance $\epsilon^2$ and zero covariance. In each interval of length $d$ we minimise the $\chi^2$ for the $n_c d$ observations made in that interval and obtain an equation of the form
\begin{equation}\label{minChi2}
    \frac{1}{\epsilon^2}\sum_{i=1}^{n_cd} \mathsf{M}^{i}\delta\mathsf{S}=\mathsf{D} \,,
\end{equation}
where the summation is over the number of observations, and we defined the vector ${\mathsf{D}}$ which contains the data. We then approximate the summation with its average times the number of observations as
\begin{equation}
    \sum_{i=1}^{n_cd}  \mathsf{M}^i \simeq \frac{n_cd}{2\pi}\int_{0}^{2\pi} \mathsf{M}^i\,\di \Theta'=n_c d\,\bar{\mathsf{M}} \,,
\end{equation}
where $\di\Theta'$ should be replaced by $\di\Psi'$ in the ELL1 model. Finally, the covariance matrix is obtained as
\begin{equation}
    \mathsf{C} = \frac{\epsilon^2}{n_cd}\bar{\mathsf{M}}^{-1} \,,
\end{equation}
from where we can read off the variances for all the parameters that define the system---we collect these variances in \autoref{app:var}.

\paragraph{Step 2: time-dependence} The second step consists of fitting the variances obtained in the first step to a time-dependent function which captures the evolution of the perturbations to the binary system. In \cite{Teukolsky1976} the time-evolution of the orbital parameters is assumed to be a polynomial up to the quadratic order. However, since we are seeking to detect the sinusoidal perturbation exerted by ULDM, we employ a Bayesian approach that mirrors the one for gravitational waves \cite{Moore:2014eua}.

We assume that the number of measurements $\mathcal{N}=T_\text{obs}/d$ of the parameter set \(\delta\mathsf{S}\) is large and that they are regularly made within the observation time. For any of the parameters \(S\) for which a variance has been estimated in step 1, $\sigma^2_{S}=\text{var}(\delta S)$ where \(\{\delta S(t)\} \deq \delta\mathsf{S}\), we define
\begin{subequations}\label{modelS}
    \begin{align}
    \delta S(t) &\deq {\bf m}(\vec{\Xi},t)+{\bf h}(t)+{\bf n}\,, \\
    {\bf h}(t) &\deq \alpha\left[{\bf A}_X(t)X+{\bf A}_Y(t)Y\right]\,, \label{heffi} \\
    {\bf m}(\vec{\Xi},t) &\deq \vec{\Xi}^T \cdot \vec{N} \quad \mbox{with}\,\,
    \vec{\Xi}^T \deq\{\Xi_0, \Xi_1,\Xi_2\} \quad \mbox{and}\,\, \vec{N}^T\deq\{1,t,t^2\}\,;
    \end{align}
\end{subequations}
here ${\bf m}$ and ${\bf h}$ are $\mathcal{N}-$dimensional vectors that describe the model---note that these vectors are \emph{not} of the same dimensionality as the parameter sets \(\delta\mathsf{S}\). The parameters $\Xi_0$, $\Xi_1$ and $\Xi_2$ are the constants that define the quadratic fit of \cite{Teukolsky1976} (note that some of the variables have $\Xi_2=0$); ${\bf n}$ is the noise of the measurement, which is assumed to be white, Gaussian and uncorrelated between pulsars; \({\bf h}\) is the signal caused by the presence of ULDM. For future convenience, here we have also separated the sinusoidal ULDM contribution into its \(A_X(t)\) and \(A_Y(t)\) contributions where $X\deq\sqrt{2}\,\varrho\cos\Upsilon$ and $Y\deq\sqrt{2}\,\varrho\sin\Upsilon$.\footnote{When $h(t)=\alpha\sqrt{2}\varrho \cos(mt+\Upsilon)$ the split is simply \( A_X(t) = \cos(mt)\) and \( A_Y(t)=- \sin(mt)\).} For instance, in the ELL1 model, from \autoref{eqaecc} we find
\begin{subequations}
\begin{align}
    \delta{x}(t) &= \delta{x}_\text{asc}+\dot{x} (t-T_\text{asc})-2 \alpha x \Phi_0 \varrho [\cos(m t+\Upsilon)-\cos(m T_\text{asc}+\Upsilon)] \nonumber\\&= \Xi_0^x + \Xi_1^x t + h^x(t) \,,\\
     h^x(t) &= -2 \alpha x \Phi_0 \varrho  [\cos(m t+\Upsilon)-\cos(m T_\text{asc}+\Upsilon)]\,,\\
     A_X(t)&=-\sqrt{2}  x \Phi_0   [\cos(m t)-\cos(m T_\text{asc})]\,,\,\,\,\,A_Y(t)=\sqrt{2}  x \Phi_0   [\sin(m t)-\sin(m T_\text{asc})]\,,
\end{align}
\end{subequations}
where  \(\Xi_2^x = 0\) in this case as there is no quadratic term---further explicit expressions for the variations \(\delta S\) can be found in \autoref{app:bayes}.

The time-model for a given orbital parameter variation needs to be tested against an alternative model wherein there is no ULDM, namely \({\bf h}(t)=0\). We call these two competing models the \emph{signal} and \emph{noise} hypotheses, respectively. In order to account for our ignorance in the quadratic fitting parameters as well as the specific realisation of the local ULDM at the location of the pulsar, we marginalise over $\Xi_j$ (which include possible secular variations of the orbital parameters) assuming uniform prior. Moreover, unless otherwise specified, we also use a uniform prior for the ULDM phase $\Upsilon$, whereas for $\varrho$ we use \autoref{Pr}; thence, the probability distribution functions for the \(X\) and \(Y\) variables is given by
\begin{equation}\label{Pxy}
    P(X,Y)=\frac{1}{2\pi}  e^{-\frac{X^2+Y^2}{2}} \,.
\end{equation} 
Lastly, the noise distribution can be written as
\begin{equation}
    P({\bf{n}}) d{\bf{n}}=\frac{\exp \left(-\frac{1}{2}{\bf{n}}^T{\bf{\Sigma_n}}^{-1}{\bf{n}}\right) d{\bf{n}}}{\sqrt{(2\pi)^\mathcal{N} \det(\Sigma_n)}}\,,
\end{equation}
where $\bf{\Sigma_n}$ is a diagonal matrix with all diagonal elements identical to $\sigma^2_{S}$, calculated in the first step.

To compare the two hypotheses, the signal \({\bf h}(t)\neq0\) and the noise \({\bf h}(t)=0\) for a given parameter \(S\), we compute the ratio of the \emph{evidences} ${\mathcal{O}_h}$ and ${\mathcal{O}_n}$ for the signal and the noise hypotheses, respectively. The evidences are defined as the likelihood marginalised over all the parameters of the model. The ratio of the evidences of the two models defines the Bayes factor $\mathcal{B}$ as:
\begin{equation}
    \mathcal{B}\deq\frac{\mathcal{O}_h}{\mathcal{O}_n}\,.
\end{equation}
Explicitly, log-likelihoods for the signal and noise hypotheses are
\begin{align}
    \log \mathcal{L}(\vec{\Xi},X,Y,\alpha,m) &\propto -\frac{1}{2}({\bf d}-{\bf m}-{\bf h})^T{\bf{\Sigma_n}}^{-1}({\bf d}-{\bf m}-{\bf h}) \,,\label{eq:ll_h}\\
    \log \mathcal{L}(\vec{\Xi}) &\propto -\frac{1}{2}({\bf d}-{\bf m})^T{\bf{\Sigma_n}}^{-1}({\bf d}-{\bf m}) \,,\label{eq:ll_n}
\end{align}
where ${\bf d}$ is the $\mathcal{N}-$dimensional vector that contains the data. Following~\cite{Moore:2014eua} we further define the matrices
\begin{equation}\label{eq:svd}
    {\bf m}\deq{\bf M} \vec{\Xi}\,,\,\,\,  {\bf M}\deq{\bf U}{\bf T}{\bf V}^\dag\,,\,\,\,  {\bf U}\deq({\bf F},{\bf G}) \,,
\end{equation}
where ${\bf M}$ is the so-called design matrix which admits a singular value decomposition into the matrices ${\bf U}$ (of dimension $\mathcal{N}\times\mathcal{N}$), ${\bf T}$ (of dimension $\mathcal{N}\times 3$) and ${\bf V}$ (of dimension $3\times 3$)---${\bf U}$ is conveniently written in terms of ${\bf F}$ and ${\bf G}$ with ${\bf G}$ an $\mathcal{N}\times \mathcal{N}-3$ matrix.\footnote{The matrix \(\mathbf{M}\) is not to be confused with the matrix \(\mathsf{M}\) defined above in \autoref{minChi2}.} From this decomposition the evidence for the noise hypothesis results in 
\begin{equation}\label{evidN}
    \mathcal{O}_n=\int  \mathcal{L}(\Xi) \,\di\vec{\Xi}=\frac{\exp \left(-\frac{1}{2}{\bf{d}}^T{\bf{G}} ({\bf{G}}^T{\bf{\Sigma_n}} {\bf{G}})^{-1}{\bf{G}}^T{\bf{d}}\right) }{\sqrt{(2\pi)^{\mathcal{N}-3} \det({\bf{G}}^T\Sigma_n {\bf{G}})}}\,.
\end{equation}

In order to obtain the evidence for the signal hypothesis we need to assume priors for $\alpha$ and the mass $m$. If we adopt a stringent detection threshold (namely, a Bayes factor of $1000$) we can assume that the posterior distributions for \(\alpha\) and \(m\) are sufficiently peaked at the true values, and that the choice of any reasonable prior for them would not affect this. Therefore, since we are interested in evaluating the sensitivity curve for $\alpha$ as a function of $m$, we use a delta-function prior for $\alpha$ and $m$ centred at their fiducial values $\alpha_f$ and $m_f$
\begin{equation}
    P(\alpha,m)=\delta(\alpha-\alpha_f) \delta(m-m_f)\,.
\end{equation}
Hence, the evidence for the signal hypothesis is given by
\begin{equation}
    \mathcal{O}_h=\int P(X,Y) P(\alpha,m) \mathcal{L}(\Xi,X,Y,\alpha,m) \,\di\vec{\Xi}\, \di X\,\di Y\,\di \alpha\, \di m\,.
\end{equation}
After performing the two trivial integrations over $\alpha$ and $m$, we are left with a number of gaussian integrals which can be computed straightforwardly, see \autoref{app:bayes}.

Finally, we average the Bayes factor over many realisations of the noise \(\bar{\mathcal{B}} \deq \int \di {\bf n} P({\bf n}) \mathcal{B}\). The result is
\begin{subequations}\label{eq:Bbar}
    \begin{align}
    \bar{\mathcal{B}}&= \exp \left\{\alpha_f^2\frac{(u_X^2+u_Y^2)}{2}\right\}\,,\\
        u_{X} &\deq {\bf A}_X^T{\bf G}({\bf G}^T{\bf\Sigma_n}{\bf G})^{-1}{\bf G}^T {\bf h}\,,\\
        u_{Y} &\deq {\bf A}_Y^T{\bf G}({\bf G}^T{\bf\Sigma_n}{\bf G})^{-1}{\bf G}^T {\bf h}\,,
    \end{align}
\end{subequations}
where in these quantities (see their definition in \autoref{app:bayes}) all the parameters are replaced by their fiducial values (i.e.\ $\alpha\to\alpha_f$, $m\to m_f$, $\varrho\to \varrho_f$, and $\Upsilon\to\Upsilon_f$). If instead of using $P(X,Y)$ given in \autoref{Pxy} we use delta-function priors also for $X$ and $Y$ (which is equivalent to assuming that we know the exact ULDM configuration for a given pulsar), we find \(\bar{\mathcal{B}} = \exp \left\{ u\right\}\) where \(u \deq {\bf h}^T{\bf G}({\bf G}^T{\bf\Sigma_n}{\bf G})^{-1}{\bf G}^T {\bf h}\).

At last, we can combine all binary pulsars by multiplying the Bayes factors as
\begin{equation}\label{eq:bayesProd}
    \bar{\mathcal{B}}^C = \prod_{p=1}^{N_p} \bar{\mathcal{B}}_p \,,
\end{equation}
where $N_p$ is the total number of binary pulsars, $\bar{\mathcal{B}}_p$ corresponds to the p-th binary system for which the Bayes factors individually are given in \autoref{eq:Bbar}. The constraint \(\alpha_f(m_f)\), which we call the \emph{sensitivity}, is obtained by choosing a threshold value for detection, customarily set to be
\begin{align}\label{eq:bayesThr}
    \bar{\mathcal{B}}^C&=\bar{\mathcal{B}}^C_\text{thr}=1000 \,,
\end{align}
see~\cite{Moore:2014eua}. For simplicity, in what follows, and in particular in \autoref{sec:res} and \autoref{sec:end}, we will drop the subscript ``f'' for the fiducial values of \(\alpha\) and \(m\).

%-------------------------------------------------------------------------------
\section{Results}
\label{sec:res}
%-------------------------------------------------------------------------------

%-------------------------------------------------------------------------------
\subsection{Near-circular orbits}\label{ssec:res_ell1}

As a first application of our method, we apply it to a set of 19~binary pulsars for which the orbital parameters were obtained by the NANOGrav collaboration~\cite{nanograv2023} using the ELL1 model and 3~binary pulsars modelled with the ELL1H (a variant on ELL1 which differs only in the parametrisation of the relativistic Shapiro time delay which we do not consider here~\cite{Hobbs:2006cd,Luo:2020ksx})---we list the properties of all the binary pulsars that we use in this work in \autoref{tab:22_ELL1psrs} and \autoref{tab:3_high_e_PSRs}. For each binary system we use the NANOGrav values of $T_\text{obs}$ (MJD), $\epsilon$ and $n_c$, and assume $\rho_\text{DM}=0.3\,{\rm GeV}/\rm{cm}^3$. Notice that, as we are forecasting the constraining power of data, and not running an actual data analysis, we need to assume the fiducial values for \(\varrho\) and \(\Upsilon\), or, equivalently, \(X\) and \(Y\); in other words, we need to simulate our data vectors and with them the unknown values of the ULDM parameters; we do this by assigning separately to each binary system a random realisation of such parameters as taken from their corresponding theoretical distributions.

\begin{figure}
    \centering
    \includegraphics[width=\textwidth]{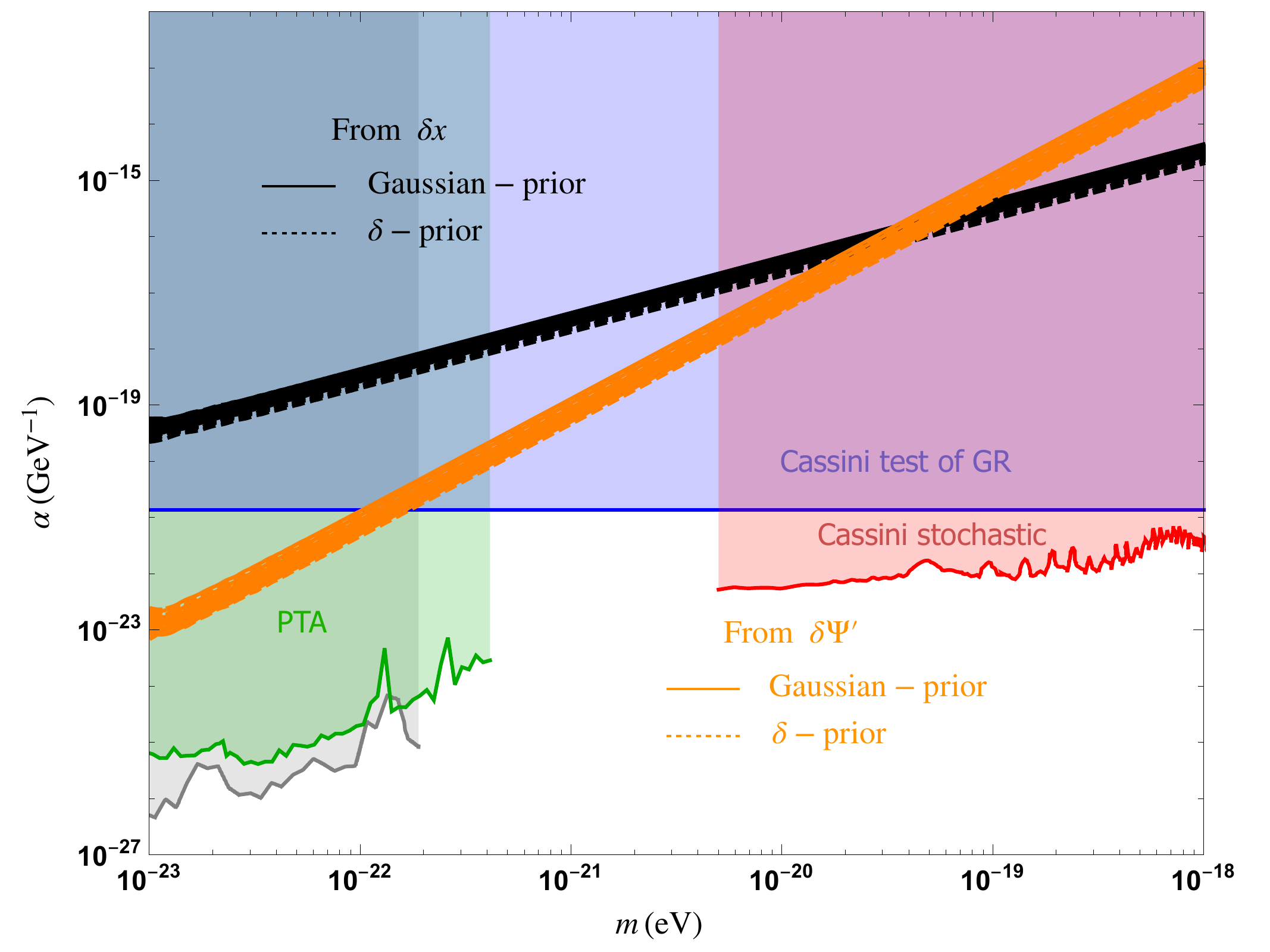}
    \caption{Sensitivity of 22 near-circular binary pulsars, as measured by NANOGrav, to the coupling $\alpha$ as a function of the ULDM mass $m$. The sensitivities obtained from \(\delta x\) are in black, whereas those pertaining to \(\delta\Psi'\) are in orange. In each case we show the results for ten different realisations of the ULDM parameters \(\varrho\) and \(\Upsilon\)---see the text for further details. We also show the constraints on \(\alpha\) coming from solar-system tests (blue), stochastic Cassini (red) and PTAs (green and grey).}
    \label{fig:nanograv_ell1}
\end{figure}

In \autoref{fig:nanograv_ell1} we show the sensitivity obtained by applying \autoref{eq:bayesThr} to the parameters \(\delta x\) and \(\delta\Psi'\) for ten different realisations of the fiducial ULDM model. In this figure we also show the current bounds obtained by solar system constraints~\cite{Bertotti:2003rm} and~\cite{Armstrong:2003ay}, and the bounds obtained with pulsar-timing arrays (PTAs)~\cite{Porayko:2018sfa,EuropeanPulsarTimingArray:2023egv}. We see that, first of all, for a wide range of ULDM masses the sensitivity of binary pulsars supersedes the current solar system constraints. This is mostly coming from the \(\delta\Psi'\) orbital parameter, which is by far more sensitive for small masses. The \(\delta x\) parameter, closely related to the semi-major axis \(a\), is the more sensitive of the two at large ULDM masses, but it currently does not give a competitive constraint. However, current PTA searches provide the most stringent limits on \(\alpha\) in the low-frequency regime, which are two to three orders of magnitude better than our limits.

\begin{figure}
    \centering
    \includegraphics[width=\textwidth]{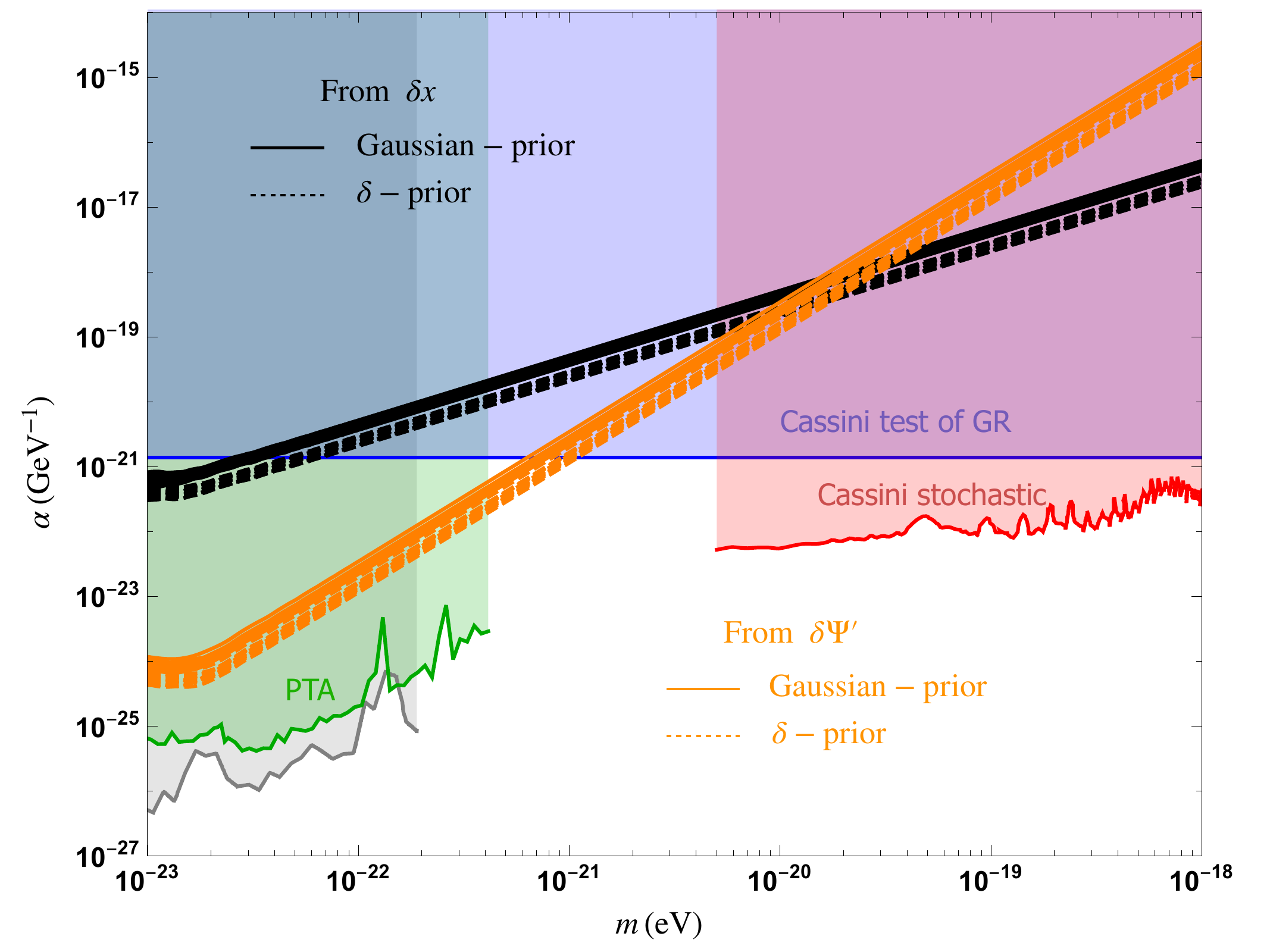}
    \caption{Same as in \autoref{fig:nanograv_ell1} but using 111 binary pulsars from the ATNF catalogue and assuming next-generation radio-telescope precision.}
    \label{fig:atnf_ell1}
\end{figure}

Thanks to the near-analyticity of our model we can readily predict how much this result would improve for any given set of parameters of future observations. In \autoref{fig:atnf_ell1} we combine 111 systems taken from the ATNF pulsar catalogue~\cite{Manchester:2004bp} and, based on the analysis presented in~\cite{Liu:2011cka} for a new generation of radio telescopes (e.g.\ the Square Kilometre Array, SKA), as an illustration of forecasting  for such experiments we assume $T_\text{obs}=10\,\text{years}$, $\epsilon=0.1\,\mu \text{s}$, $n_c=1/\,\text{day}$. As we can see in the figure, the sensitivity in this case improves by well over an order of magnitude, bringing the binary pulsar constraint below the solar system one for a larger range of ULDM masses. In particular, the sensitivity obtained with our method extends to masses beyond the Nyquist frequency of PTAs, and can beat solar system tests in that range by up to one order of magnitude.

In order to show how our results depend on the choice of priors for the ULDM parameters \(X\) and \(Y\) (corresponding to choices of \(\varrho\) and \(\Upsilon\)), in \autoref{fig:nanograv_ell1} and \autoref{fig:atnf_ell1} we also show the sensitivity curves for delta-function priors \(\delta(X_\text{fid}-X)\) and \(\delta(Y_\text{fid}-Y)\). We observe that, while the choice of prior is not entirely negligible, nonetheless the effect is limited to a factor of order unity. Lastly, notice that the variations of the Laplace-Lagrange orbital parameters \(\eta\) and \(\kappa\) do exhibit resonances (see \autoref{deltaKappa} and \autoref{deltaEta}); however, beyond the resonant peak these parameters are less sensitive than \(x\) and \(\Psi'\).

%-------------------------------------------------------------------------------
\subsection{Resonances}\label{ssec:res_bt}

\begin{figure}
    \centering
    \includegraphics[width=\textwidth]{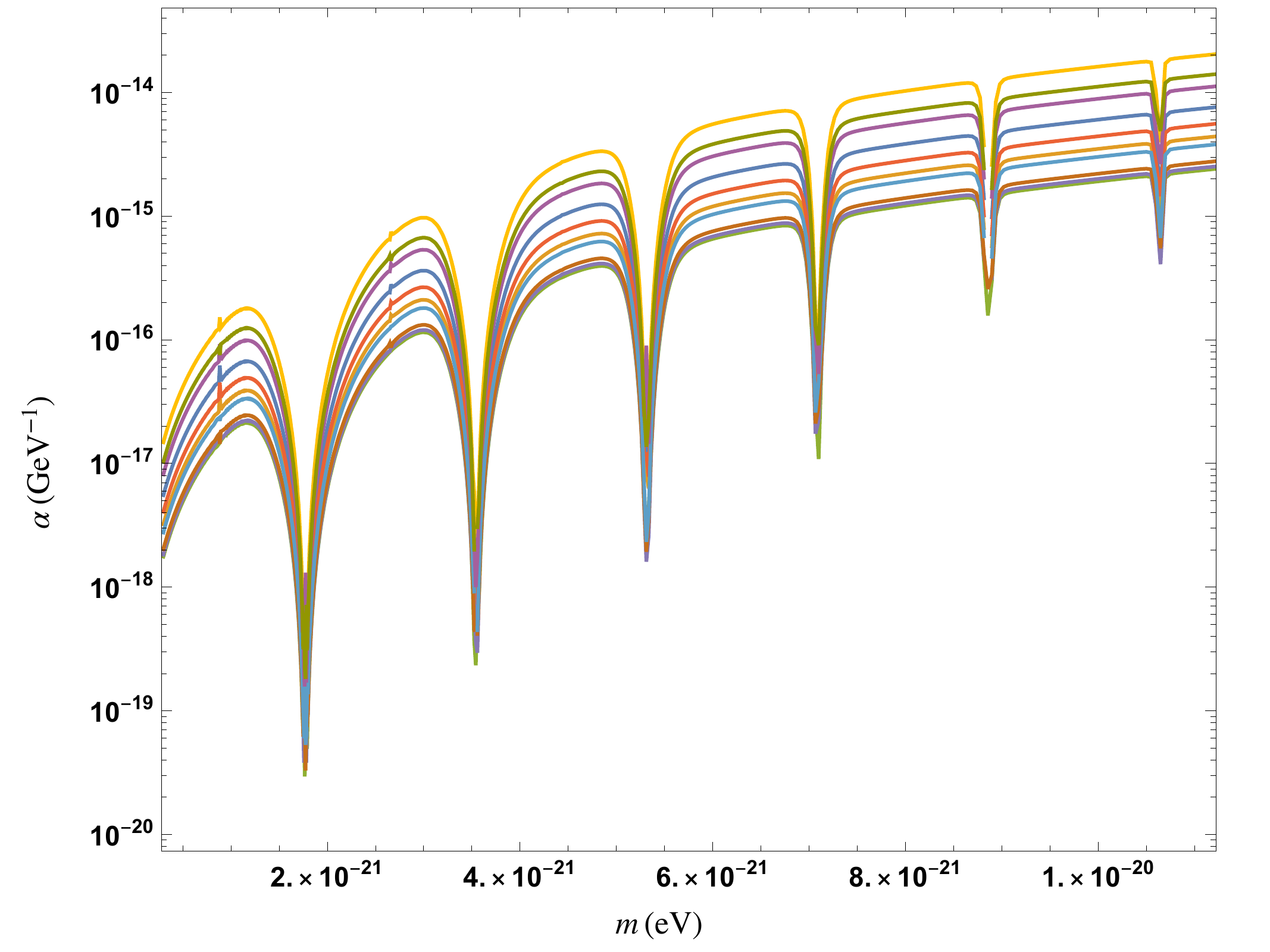}
    \caption{Sensitivity to $\alpha$ as a function of ULDM mass $m$ for ten realisations of the ULDM parameters $\varrho$ and $\Upsilon$, obtained from the binary pulsar J1946+3417 as measured by NANOGrav.}
    \label{fig:nanograv_bt1}
\end{figure}

\begin{figure}
    \centering
    \includegraphics[width=\textwidth]{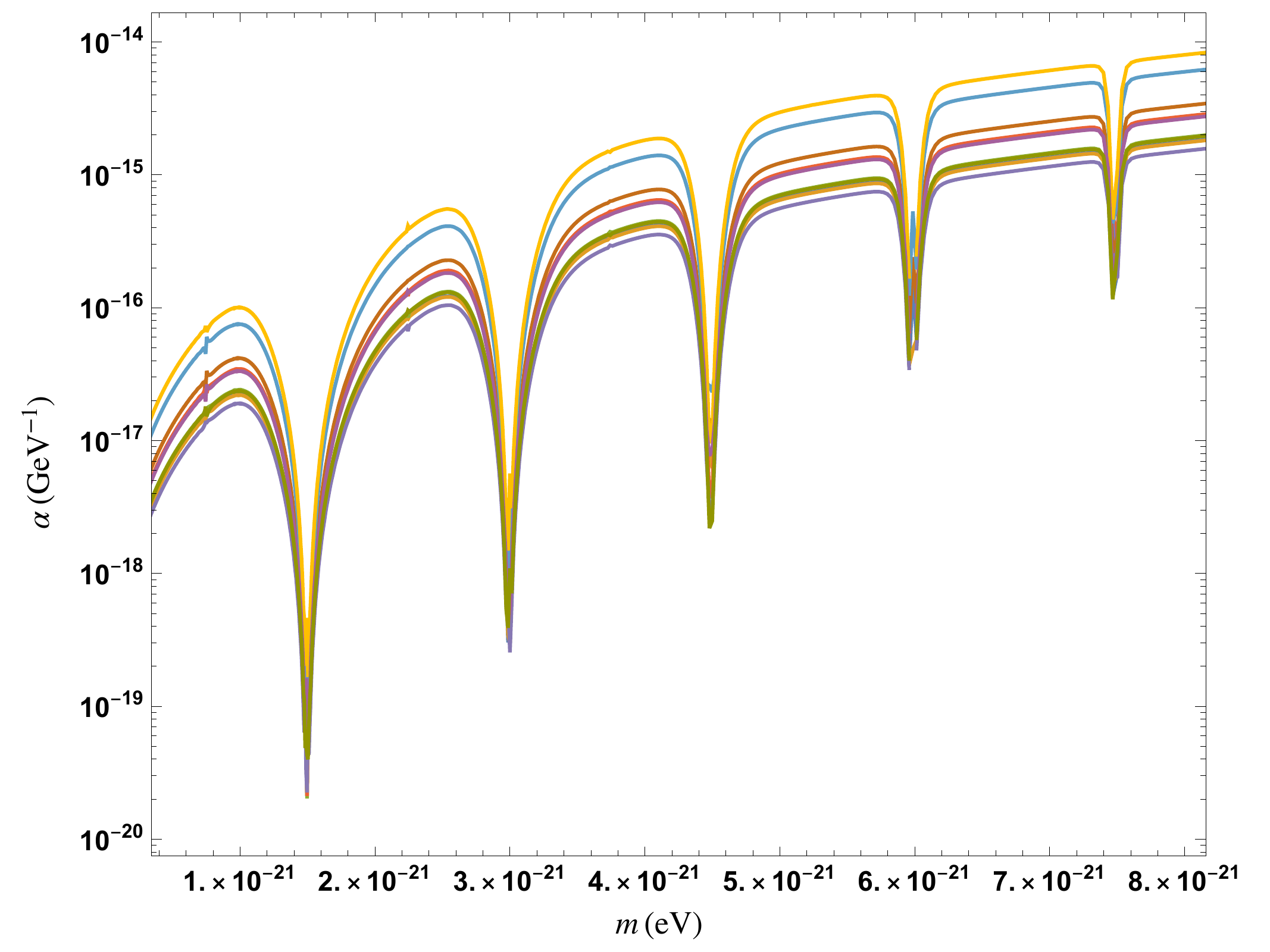}
    \caption{Sensitivity to $\alpha$ as a function of ULDM mass $m$ for ten realisations of the ULDM parameters $\varrho$ and $\Upsilon$, obtained from the binary pulsar J2234+0611 as measured by NANOGrav.}
    \label{fig:nanograv_bt2}
\end{figure}

\begin{figure}
    \centering
    \includegraphics[width=\textwidth]{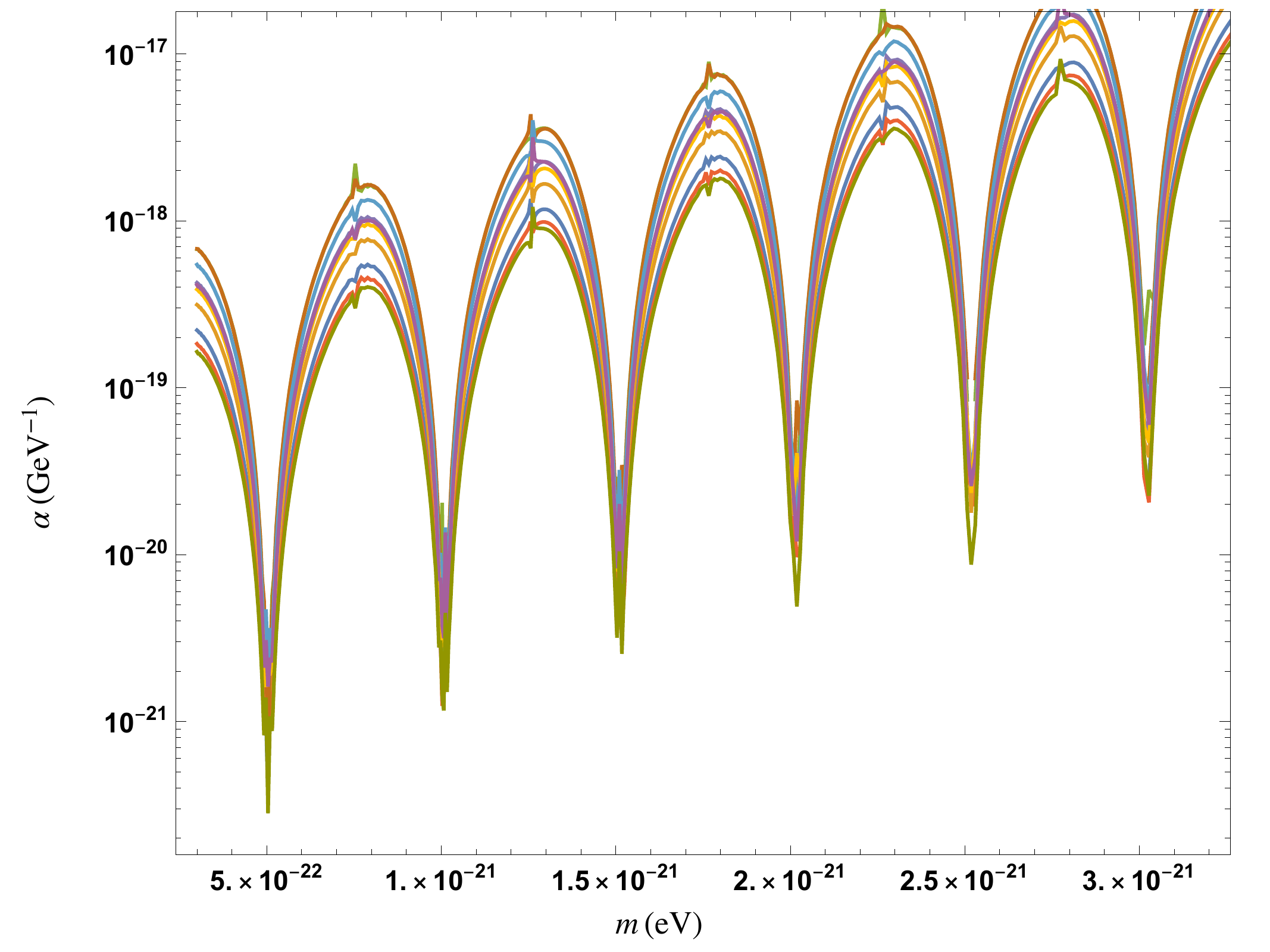}
    \caption{Sensitivity to $\alpha$ as a function of ULDM mass $m$ for ten realisations of the ULDM parameters $\varrho$ and $\Upsilon$, obtained from the binary pulsar J1903+0327 as measured by NANOGrav.}
    \label{fig:nanograv_bt3}
\end{figure}

When a system possesses a significant eccentricity \(e\) its orbital parameters experience resonant secular drifts~\cite{Blas:2016ddr,Blas:2019hxz}, and obtaining the sensitivity becomes much more computationally intensive. Therefore, here we limit ourselves to the effects of ULDM on \(P_b\), and not the full variation of the parameter \(\Theta'\) defined in \autoref{eq:defThp} and \autoref{eq:varThp} and refer to \autoref{app:beyondPb} for a more detailed discussion on this simplification, which does not appreciably alter our results. We call this contribution $\delta\Theta'_{P_b}$, defined as
\begin{eqnarray}\label{eq:varThpPb}
    \delta\Theta'_{P_b} &=&\delta\Theta'_0-\int_{T_0}^{t}{\omega_b}\frac{\delta P_b}{P_b} \di t'\,. 
\end{eqnarray}
With this proviso, in \autoref{fig:nanograv_bt1}, \autoref{fig:nanograv_bt2} and \autoref{fig:nanograv_bt3} we show the sensitivity obtained from this simplified \(\delta\Theta'\) for the three binary pulsars J1946+3417, J2234+0611 and J1903+0327, respectively (we collect their parameters in \autoref{tab:3_high_e_PSRs}); as in \autoref{ssec:res_ell1} we run ten realisations of the ULDM parameters for each binary system.

\begin{figure}
    \centering
    \includegraphics[width=\textwidth]{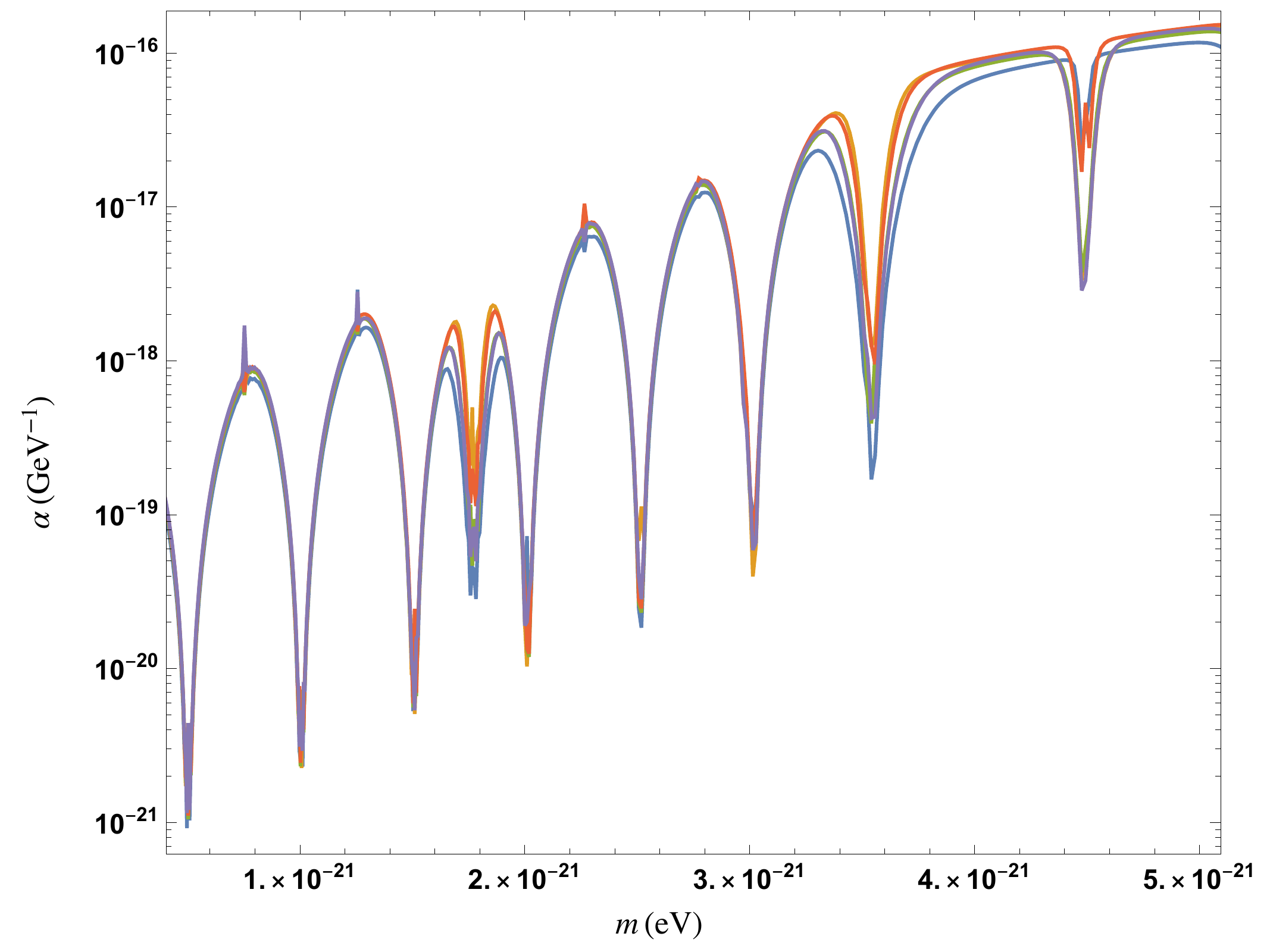}
    \caption{Sensitivity to $\alpha$ as a function of ULDM mass $m$ for five realisations of the ULDM parameters $\varrho$ and $\Upsilon$, obtained from the combination of the three binary pulsars J1946+3417, J2234+0611 and J1903+0327 as measured by NANOGrav.}
    \label{fig:nanograv_bt}
\end{figure}

The combined sensitivity from all three systems, which beats each of them separately, as expected, is shown in \autoref{fig:nanograv_bt}. In these figures we limit ourselves to a smaller mass range to focus on the resonances, and we can see how these systems are in effect less constraining than the near-circular ones, their combined sensitivity approaching that of solar system tests only for the most peaked resonances at low masses. Finally, the results of this section are directly comparable to~\cite{Blas:2019hxz}; we see that, while we reproduce these results very closely near the resonances, the off-resonance sensitivity as obtained from our method is in fact better than originally estimated.

%-------------------------------------------------------------------------------
\section{Conclusion and outlook}
\label{sec:end}
%-------------------------------------------------------------------------------

\paragraph*{Summary} In this paper we computed the sensitivity of binary pulsars for the detection of scalar ULDM for the entire mass range \(10^{-23}~\mathrm{eV}\lesssim m \lesssim10^{-18}~\mathrm{eV}\), and for both eccentric and near-circular systems. We improve on previous results which only considered resonances between the systems' orbital periods and ULDM oscillations by developing a method that goes beyond resonances. With the new method it is also possible to quantify our ignorance on the local DM field (i.e.\ the local amplitude determined by the parameter $\varrho$ and the phase $\Upsilon$). We find that, while our method is not competitive in the range of frequencies surveyed by PTAs, beyond the Nyquist frequency binary pulsars can be the most sensitive detectors for ULDM. For instance, we expect that observations made with the future SKA would be sensitive to ULDM couplings \(\alpha\approx\text{few}\times 10^{-22}/\text{GeV}\) for \(m\approx\text{few}\times 10^{-22}\,\text{eV}\), better than solar system constraints. In this regard, we notice that the most sensitive orbital parameter is \(\Psi'\) (for ELL1 binaries) or \(\Theta'\) (for BT binaries), highlighting the relevance of orbital parameters other than the most-studied (and often best-measured) orbital period \(P_b(t)\).

\paragraph*{Relevance} Our results highlight why it is important to go beyond resonances. Firstly, resonances, being narrow resonances, are limited to a very narrow range of ULDM masses in correspondence with integer multiples of the binary system inverse period: going beyond resonances allows us to extend the sensitivity of each binary system to the entire mass range of interest, only limited by the binary pulsar back-reactions on the field \(\Phi\) on the higher end~\cite{Blas:2019hxz} and cosmological constraints on \(\rho_\text{DM}\) on the lower end (see, e.g.\ \cite{Shevchuk:2023ccb} and references therein). Secondly, in the scalar ULDM case resonances disappear for near-circular orbits; since near-circular orbits constitute the majority of binary pulsars in nature, for scalar ULDM searches we are limited to a minority of systems. Going beyond resonances means that each observed binary system contributes to the sensitivity for ULDM. Thirdly, each binary system contributes separately to the sensitivity to ULDM, as obtained from the secular orbital drifts caused by resonances. With our new method all binary pulsars jointly contribute to the sensitivity, thereby improving it across all masses.

\paragraph*{Context} Our findings should be compared with existing limits on scalar ULDM. We have already shown the limit coming from the Doppler tracking of the Cassini spacecraft, which results in \(\alpha\lesssim10^{-21}\,\text{GeV}^{-1}\) across all the masses we consider here~\cite{Bertotti:2003rm}. Because a putative ULDM direct coupling \(\alpha\) is equivalent to a scalar gravitational wave, the Cassini spacecraft also sets a constrain on the mass range \(5\times10^{-21}~\mathrm{eV}\lesssim m \lesssim10^{-18}~\mathrm{eV}\) that is about an order of magnitude stronger~\cite{Armstrong:2003ay}. In the lowest frequency range considered in this work, current searches of the stochastic gravitational wave background by pulsar-timing arrays set the limits on ULDM around \(\alpha\lesssim10^{-26}\,\text{GeV}^{-1}\) for a decade in mass around \(10^{-23}~\mathrm{eV} \lesssim m \lesssim 10^{-22}~\mathrm{eV}\) \cite{Porayko:2018sfa,EuropeanPulsarTimingArray:2023egv}. Furthermore the dynamical friction within the solitonic core of the ULDM halo changes the spectral slope of the stochastic background of gravitational waves which is at odds with what has been measured by pulsar-timing arrays, possibly implying the exclusion of ULDM masses within the range \(1.3\times10^{-21}~\mathrm{eV}\lesssim m \lesssim1.4\times10^{-20}~\mathrm{eV}\) \cite{Aghaie:2023lan}. Existing limits also come from superradiance. By measuring the spin and mass of known black holes and other astrophysical objects, and assuming no interaction between \(\Phi\) and standard matter, the mass ranges \(2.5\times10^{-21}~\mathrm{eV}\lesssim m \lesssim1.2\times10^{-20}~\mathrm{eV}\) and \(5.5\times10^{-20}~\mathrm{eV}\lesssim m \lesssim1.3\times10^{-16}~\mathrm{eV}\) are excluded, or else these black holes and celestial objects would not be there~\cite{Stott:2020gjj}. Therefore, the limits we obtain from binary pulsars at the same time constitute an independent test of some of the parameter space that is probed by superradiance, as well as a test of regions not accessible by it.

\paragraph*{Prospects} In this work we have concerned ourselves with the simplest case of a universally-coupled linear scalar ULDM field only. In the future we plan to investigate the spin-1 and spin-2 ULDM cases, for which only the resonant effects have been studied~\cite{LopezNacir:2018epg,Armaleo:2019gil}. The main difference with respect to the scalar case is that, if ULDM has spin 1 or spin 2, even circular orbits experience resonant secular orbital drifts. Moreover, since the geometry of the ULDM perturbations on the orbital motion is different, we expect that it is possible to differentiate between different ULDM spins. The spin-2 case is especially interesting because ULDM with spin 2 mimics a continuous gravitational wave, albeit with five polarisations, see~\cite{Armaleo:2020efr}. Important extension to this work are the cases of a non-universal coupling as well as a quadratic coupling, which present distinctive phenomenology compared to the linear case, see, e.g.\ \cite{Blas:2019hxz}. Furthermore, a more detailed analysis would include other sources of noise (such as pulsar red noise) as well as more complex and complete timing models, which are optimally tailored to each binary pulsar system. Finally, we plan to empirically test our results by directly injecting a model ULDM signal to simulated binary pulsar time-of-arrival data, and then using different methods (applied to simulated data) in order to recover the injected ULDM signal within the predicted level of sensitivity.

With the advent of the next generation of radio surveys such as the SKA we expect to observe an order-of-magnitude more binary pulsars than today. With our method it is immediate to combine all the systems, each with very different properties, into a single sensitivity curve across all relevant masses, and thus assess the chances for either ULDM detection, or the tightest constraints on its parameter space as new systems are discovered.

\paragraph*{Acknowledgements}
The Authors wish to thank Stéphane Ili\'c for useful discussion. DLN acknowledges support from ESIF and MEYS (Project ``FZU researchers, technical and administrative staff mobility'' - CZ.02.2.69/0.0/0.0/18\_053/0016627),  UBA and CONICET. PK and FU acknowledge support from MEYS through the INTER-EXCELLENCE II, INTER-COST grant LUC23115.

%-------------------------------------------------------------------------------
\appendix
%-------------------------------------------------------------------------------

%-------------------------------------------------------------------------------
\section{Post-Keplerian orbits}\label{app:sec}

The orbital parameters of a perturbed binary system evolve with time according to the equations~\cite{Turner:1979yn,Danby:1970},
{\allowdisplaybreaks
\begin{subequations}\label{Lagrange}
\begin{align}
\label{Lagrange1}
\dot a &=
\frac{2}{\omega_b}\left\{\frac{F_{r}
    e}{\sqrt{1-e^2}}\sin \theta+\frac{F_{\theta}}{r}a\sqrt{1-e^2}\right\}\,, \,\,\,\\
\label{Lagrange2}
\dot e&= \frac{\sqrt{1-e^2}}{a\omega_b}\left\{F_{r} \sin
  \theta+F_{\theta}\bigg(\cos 
  \theta+\frac{1}{e}-\frac{r}{ae}\bigg)\right\}\,, \\
\label{Lagrange3}
\dot \Omega&= \frac{F_{z}r
  \sin(\theta+\omega)}{a^2\omega_b\sqrt{1-e^2}\sin\iota}\,, \\
\label{Lagrange4}
\dot\iota&=\frac{F_{z}r \cos(\theta+\omega)}{a^2\omega_b\sqrt{1-e^2}}\,, \\
\dot \varpi&= \frac{\sqrt{1-e^2}}{a e\omega_b}
\left\{-F_{r} \cos \theta+F_{\theta}\sin \theta\left[1+\frac{r}{a(1-e^2)}\right]\right\} + 2\sin^2\left(\frac{\iota}{2}\right)\dot\Omega\,,
\label{Lagrange5}  \\
\dot \epsilon_1&=
-\frac{2rF_{r}}{a^2\omega_b}+\left[1\!-\!\sqrt{1-e^2}\right]
\dot\varpi + 2\sqrt{1-e^2}  \sin^2\left(\frac{\iota}{2}\right)
\dot\Omega\,,
\label{Lagrange6}
\end{align}
\end{subequations}
}
where
\begin{equation}
\label{orbfreq}
\omega_b=\sqrt{\frac{GM_T}{a^3}},
\end{equation}
is the binary orbital frequency and we have decomposed the perturbing force in the cylindric coordinates,
\begin{equation}
\label{Fdecomp}
\vec{F}=F_r\hat{r}+F_{\theta}\hat{\theta}+F_z\hat{z}\;.
\end{equation}
In \autoref{Lagrange5} and \autoref{Lagrange6} we have introduced the argument of the pericentre, 
\begin{equation}
\varpi=\omega+\Omega\,,
\end{equation}
and the parameter
\begin{equation}\label{eps1_2}
\epsilon_1=\omega_b(t-T_0)+\varpi-\int_{T_0}^t \di t' \,\omega_b\,.
\end{equation} 

The orbital motion for elliptic orbits is not expressible in closed form, but can be represented as Fourier decomposition of combinations of the unperturbed Keplerian orbits, which amounts to a harmonic expansion in eccentricities~\cite{watson1995treatise}:
{\allowdisplaybreaks
\begin{subequations} 
\label{KeplFour}
\begin{align}
& \frac{r}{a}= 1+\frac{e^2}{2}-2e\sum_{n=1}^{\infty}
\frac{J_n'(ne)}{n}  \,\cos n\omega_b \tilde t\,,\\ 
& \frac{r^2}{a^2}= 1+{ \frac{3}{2}}e^2-4\sum_{n=1}^{\infty}
\frac{J_n(ne)}{n^2}  \,\cos n\omega_b \tilde t\,,\\ 
&\cos \theta=-e+\frac{2(1-e^2)}{e} \sum_{n=1}^{\infty} J_n(ne) \,
\cos n\omega_b \tilde t\,,\\
&\sin \theta= 2\sqrt{1-e^2}\sum_{n=1}^{\infty}  J_n'(ne) \,\sin
n\omega_b \tilde t\,,\\
&\frac{r}{a}\cos \theta=-\frac{3e}{2}+2
\sum_{n=1}^{\infty} \frac{J_n'(ne)}{n} \,\cos n\omega_b \tilde t\,,\\ 
&\frac{r}{a}\sin \theta=\frac{2\sqrt{1-e^2}}{e}
\sum_{n=1}^{\infty} \frac{J_n(ne)}{n} \,\sin n\omega_b \tilde t\,,\\ 
&\frac{a^2}{r^2}\cos \theta= \sum_{n=1}^{\infty}2n J_n'(ne) \,\cos
n\omega_b \tilde t\,,\\ 
&\frac{a^2}{r^2}\sin \theta=
\frac{\sqrt{1-e^2}}{e}\sum_{n=1}^{\infty}2n  J_n(ne) \,\sin n\omega_b
\tilde t\,, 
\end{align}
\end{subequations}
}
where $\tilde t=t-T_0$,  $J_n(z)$ is a Bessel function of order $n$ and $J_n'(z)$ is its derivative with respect to $z$.

%-------------------------------------------------------------------------------
\section{Variances}\label{app:var}

\paragraph*{Elliptic orbits}
The time residual for the BT model is given by \autoref{ResBT} in terms of six independent parameters: \(\{\delta K\,,\delta a\,,\delta e\,,\delta \alpha_b\,,\delta \eta_b\,,\delta\Theta'\}\). The dependence on $\delta a$  only comes from the dependence on $\omega_b$ in the last term of \autoref{Npulset},  which is a subdominant contribution with respect to the other terms. Here, for simplicity, we   follow  \cite{Teukolsky1976} and we neglect  such contribution  to estimate the variances of the remaining five parameters: \(  \{\delta K\,,\delta e\,,\delta \alpha_b\,,\delta \eta_b\,,\delta\Theta'\}\). The dominant part of the time residual can be written as 
\begin{align}\label{ResBT3}
R^{\mathrm{BT}} =&\,
\delta \eta_b \sin  E'- \delta e \left[\alpha_b +\frac{\sin E'  \left(\alpha_b \sin E'-\eta_b \cos E'\right)}{1-e \cos E'}\right]+\delta \alpha_b \left(\cos E'-e\right)\nonumber\\
&-\frac{\delta \Theta ' \left(\alpha_b  \sin E'-\eta_b  \cos E'\right)}{1-e \cos E'} \,,
\end{align} and the variances for the parameters for generic values of $e$, $\alpha_b$ and $\eta_b$ are as follows.
{\allowdisplaybreaks
\begin{subequations}
\begin{align}
    \text{var}({\delta K})=&-\mathcal{Q} \left\{ \eta_b^4 \left[e^4 (26-9 \te)+4 e^2 (8 \te-7)+8 (\te-1)\right] \right. \nonumber\\
    &+2 \alpha_b^2 \eta_b^2 \left[e^6+e^4 (18-5\te)+4 e^2 (6 \te-5) +8 (\te-1)\right] \nonumber\\
    &+\alpha_b^4 \left[8 e^8+e^6 (32 \te-62)+e^4 (74-65 \te)+4 e^2 (4\te-3)+8 (\te-1)\right]\left.\right\} \,,\\
    \text{var}({\delta \alpha_b})=&\,\mathcal{Q} \left\{ 4\eta_b^4 \left[e^2 (\te-3)-4 \te+4\right] +2 \alpha_b^2 \eta_b^2 \left[7 e^4-22 e^2-2 \left(e^4-7 e^2+8\right) \te+16\right] \right. \nonumber\\
    &+\alpha_b^4 \left[e^6 (\te-6)+e^4 (30-16 \te)+8 e^2 (4 \te-5)-16 (\te-1)\right] \left.\right\} \,,\\
    \text{var}({\delta \eta_b})=&\,\mathcal{Q} \left\{ 4\alpha_b^4 \left[e^2-1\right)^2 \left(e^2 (\te-3)-4 \te+4\right] +2 \eta_b^4 \left[e^4+4 e^2 (\te-2)-8\te+8\right] \right. \nonumber\\
    &+\alpha_b^2 \eta_b^2 \left[e^6 (\te-6)+e^4 (42-20 \te)+e^2 (52 \te-68)-32 (\te-1)\right] \left.\right\} \,,\\
    \text{var}({\delta e})=&\,\mathcal{Q} e^4 \left\{\alpha_b^2 \left(e^2-2\right) \left[e^2 (\te-2)-2 \te+2\right]-2 \eta_b^2 \left(e^2+2 \te-2\right)\right\} \,,\\
    \text{var}({\delta \Theta'})=&\,\mathcal{Q} \te e^2 \left\{2 \alpha_b^2 \left(e^2-1\right) \left[e^2 (\te-2)-2 \te+2\right]+\eta_b^2 \left(e^2-2\right) \left(e^2+2 \te-2\right)\right\} \,,\label{VarThetaprime}
\end{align}\end{subequations}
}
where
\begin{align}
    \mathcal{Q} &\deq \frac{\epsilon^2}{d n_c \left[e^4+4 e^2 (\te-2)-8 \te+8\right] \left[\eta_b^2-\alpha_b^2 \left(e^2-1\right)\right]^2} \,,
\end{align}
and $\te \deq \sqrt{1-e^2}$.

\paragraph*{Near-circular orbits} In the ELL1 timing model it is not difficult to obtain an analytic estimate for the variance of  the six orbital parameters. These expressions are however very long and not illuminating, and for simplicity here we only quote the results for $\eta\simeq\kappa\simeq0$.
{\allowdisplaybreaks
\begin{subequations}
    \begin{align}
    \text{var}({ \delta K})=&\frac{\epsilon^2 \left(34 x^4  \omega_b^4+61 x^2  \omega_b^2+76\right)}{d n_c \left(34 x^4  \omega_b^4-11 x^2
    \omega_b^2+4\right)}\,,\\
 \text{var}({ \delta s})=&   \frac{4 s^2 \epsilon^2 \left(5 x^4 \omega_b^4+80 x^2 \omega_b^2+32\right)}{81 d n_c x^6 \omega_b^4}\,,\\
   \text{var}({ \delta x})=&  \frac{20 \epsilon^2}{9 d n_c}\,,\\
\text{var}({ \delta \eta})=& \frac{32 \epsilon^2 \left(x^2  \omega_b^2+1\right)}{d n_c x^2\left(34 x^4  \omega_b^4-11 x^2  \omega_b^2+4\right)}\,,\\
   \text{var}({ \delta \kappa})=&\frac{32 \epsilon^2}{9 d n_c x^4 \omega_b^2} \,,\\
   \text{var}({ \delta \Psi'})=& \frac{4 \epsilon^2 \left(17 x^2  \omega_b^2+2\right)}{d n_c x^2 \left(34 x^4  \omega_b^4-11 x^2 \omega_b^2+4\right)}\,.
    \end{align}
\end{subequations}
}
These expressions can be further simplified by noticing that the quantity
\begin{equation}
    \omega_b^2 x^2=\frac{GM_2}{a}\frac{\sin^2\iota  }{\left(1+\frac{M_1}{M_2}\right)}
\end{equation}
is very small, so that at leading order we have:
\begin{eqnarray}\label{LOvariance}
    &&
    \text{var}({ \delta K})= \frac{19 \epsilon^2}{d n_c}\,,~~ \text{var}({ \delta s})= \frac{128 s^2 \epsilon^2}{81 d n_c x^6 \omega_b^4}\,,~~ \text{var}({ \delta x})=  \frac{20\epsilon^2}{9 d n_c}\,, \nonumber\\
    &&
    \text{var}({ \delta \eta})= \frac{8\epsilon^2}{d n_c x^2}\,,~~ \text{var}({ \delta \kappa})=\frac{32 \epsilon^2}{9 d n_c x^4 \omega_b^2} \,,~~\text{var}({ \delta \Psi'})= \frac{2 \epsilon^2 }{d n_c x^2}\,.
\end{eqnarray}
Notice that the variances of $\delta s$ and $\delta \kappa$ are  relatively large since powers of the small quantity $x\omega_b$ appear in the denominator. This is expected because the dependence on $s$ only enters in the factor $\omega_b$ in the last term of \autoref{NpulsetELL}, which (as in the BT model) is a subdominant contribution. By ignoring such contribution the calculation of the variances of the remaining five independent parameters yields:
\begin{eqnarray}\label{LOvarianceS}
    &&
    \text{var}({ \delta K})= \frac{(19+ 18\kappa^2)\epsilon^2}{d n_c}\,,~~   \text{var}({ \delta x})=  \frac{2 \epsilon^2}{ d n_c}\,, \nonumber\\
    &&
    \text{var}({ \delta \eta})= \frac{2\epsilon^2(4+\eta^2+4\kappa^2)}{d n_c x^2}\,,~~ \text{var}({ \delta \kappa})=\frac{2 \epsilon^2(4+4\eta^2+\kappa^2)}{ d n_c x^2   } \,,~~\text{var}({ \delta \Psi'})= \frac{2 \epsilon^2 }{d n_c x^2}\,.
\end{eqnarray}
Therefore, except for $\delta\kappa$, this relatively simpler calculation gives similar results. The reason why the variance of $\delta\kappa$ becomes smaller by neglecting the contribution of $\delta s$ is that at leading (non-vanishing) order in $x \omega_b$ the term with $\delta s$, which is suppressed by $x \omega_b$, becomes completely degenerate with the one with  $\delta\kappa$ (they have the same time dependence given by $\sin 2\Psi'$), which does not vanish for $x \omega_b\to 0$. It is only after including the \({\cal O}(1)\) correction in $x \omega_b$ for $\delta\kappa$ that those parameters can be distinguished. 

%-------------------------------------------------------------------------------
\section{Bayesian approach}\label{app:bayes}

\paragraph*{Perturbed BT model} The model for the variations of the orbital parameters for eccentric orbits including the ULDM contribution is given by:
{\allowdisplaybreaks
\begin{subequations}\label{signalsBT}
\begin{align}\label{eq:varThpAp}
    \delta\Theta'(t)=&\delta\Theta'_0+ \frac{5}{2}\alpha \omega_b\frac{\Phi_0\varrho}{m}[\sin(mt+\Upsilon)-\sin(m T_0+\Upsilon)]+\alpha \Phi_0\varrho \sum_{n=1}^{+\infty}   B_n \sin (n\omega_b\tilde t)\cos(m t+\Upsilon) \,\nonumber \\
    &+\alpha\frac{\Phi_0\varrho}{2}\sum_{n=1}^{+\infty} (A_n-n\omega_b B_n)\left\{\frac{ \sin(m t+\Upsilon+n\omega_b\tilde t)-\sin(m T_0+\Upsilon) }{m+n\omega_b}\right.\nonumber \\
    &\left.+\frac{ \sin(m t+\Upsilon-n\omega_b\tilde t)-\sin(m T_0+\Upsilon) }{m-n\omega_b}\right\}
    -\frac{3}{2}\omega_b\int_{T_0}^t\frac{\delta a(t')}{a} dt'\,,\\ \label{eq:vardeltaaAp} 
    \frac{\delta a(t)}{a}=&\frac{\delta a_0}{a}+\frac{\dot a_0\tilde t}{a}-2\alpha \Phi_0 \varrho  [\cos(m t+ \Upsilon)-\cos(m T_0+ \Upsilon)]\nonumber\\
    &-8 \alpha \Phi_0 \varrho \sum_{n=1}^{+\infty} J_n(ne) [\cos(mt+\Upsilon)\cos(n\omega_b\tilde t)-\cos(m T_0+\Upsilon)]\nonumber\\
    &+6 \alpha \Phi_0 \varrho \omega_b \sum_{n=1}^{+\infty} n J_n(ne)\left\{\frac{\cos(m t+\Upsilon+n\omega_b\tilde t)-\cos(m T_0+\Upsilon)}{m+n\omega_b }\right.\nonumber\\
    &\left.-  \frac{\cos(m t+\Upsilon-n\omega_b\tilde t)-\cos(m T_0+\Upsilon)}{m-n\omega_b } \right\}\,,\\
    \delta e(t)=&\delta e_0+\dot e_0\tilde t+\alpha\Phi_0 \varrho \frac{1-e^2}{e}\sum_{n=1}^{+\infty} J_n(ne) \left\{ \frac{n\omega_b-2m}{ n\omega_b+m}[\cos(m t+\Upsilon+n\omega_b \tilde t)-\cos(m T_0+\Upsilon)]\right.\nonumber\\
    &\left.+\frac{2m+n\omega_b}{n\omega_b-m}[\cos(m t+\Upsilon-n\omega_b \tilde t)-\cos(m T_0+\Upsilon)]\right\}\, ,\\
    \delta \omega(t)=&\delta \omega_0+\dot\omega_0\tilde t+\alpha \Phi_0 \varrho\frac{\sqrt{1-e^2}}{e} \sum_{n=1}^{+\infty}J_n'(ne) \left\{ \frac{n\omega_b-2m}{ n\omega_b+m}[ \sin(m t+\Upsilon+n\omega_b \tilde t)-\sin(m T_0+\Upsilon)] \right.\nonumber\\
    &\left.-\frac{2m+n\omega_b}{n\omega_b-m} [\sin(m t+\Upsilon-n\omega_b \tilde t)-\sin(m T_0+\Upsilon)] \right\}\,,
\end{align}
\end{subequations}
}where $\tilde t=t-T_0$, $A_n$ and $B_n$ are defined in \autoref{AnBn}, \(\delta a_0\), \(\delta e_0\) and \(\delta \omega_0\) account for a constant shift due to the error in the determination of the parameters evaluated at $T_0$, and the derivatives of the orbital parameters on the right-hand side account for the possibility of having another perturbation to the system (on top of the ULDM effect) which produces a secular variation. The variations \(\delta\Omega\) and \(\delta s\) do not have contributions from ULDM. Notice that in the two-step approach presented in the main text we marginalise over those errors for each variable; we can therefore discard ULDM-dependent terms that can be absorbed in the errors of the orbital parameters, since the ULDM contribution cannot be determined separately.

For the sake of completeness we write below the 
ULDM contribution decomposed into its $X$ and $Y$ components:
\begin{subequations}\label{hsignal}
\begin{align}
\delta S^{\text{DM}}(t)&=h^{S}(t)+\ell^{S}(t)\,,\\
h^S(t)&\deq \alpha \left[ A_{X}^S(t)X+A_{Y}^S(t) Y\right]\,,\\ 
A_{X}^S(t)&=   H_{X}^S(t)- H_{X}^S(T_\text{ref})\,,\\
A_{Y}^S(t)&=   H_{Y}^S(t)- H_{Y}^S(T_\text{ref})\,,\\
\ell^S(t)&\deq  \alpha \left[L_{X}^S(t)X+L_{Y}^S(t)Y\right]\,,
\end{align}
\end{subequations}
where $h^S(t)$ is the characteristic signals caused by ULDM on the variables $\delta S$ that enter the likelihood in \autoref{eq:ll_h}, for the BT model $T_\text{ref}=T_0$,
{\allowdisplaybreaks
\begin{subequations}\label{hthetaXY}
\begin{align}
     H^{\Theta'}_X(t)=& \sqrt{2}  \omega_b\frac{\Phi_0}{m}\sin mt + \frac{\Phi_0}{\sqrt{2}} \sum_{n=1}^{+\infty}   B_n \sin (n\omega_b\tilde t)\cos m t  \,\nonumber +\frac{\Phi_0}{2^{3/2}}\sum_{n=1}^{+\infty} (A_n-n\omega_b B_n)\\
    &\times\left\{\frac{ \sin(m t+n\omega_b\tilde t) }{m+n\omega_b} +\frac{ \sin(m t-n\omega_b\tilde t) }{m-n\omega_b}\right\}
    +H^{\Theta'}_{P_b,X}(t)\,,\\  
      H^{\Theta'}_Y(t)=&  \sqrt{2}  \omega_b\frac{\Phi_0}{m}\cos m t -\frac{\Phi_0}{\sqrt{2}} \sum_{n=1}^{+\infty}   B_n \sin (n\omega_b\tilde t)\sin m t \,\nonumber + \frac{\Phi_0}{2^{3/2}}\sum_{n=1}^{+\infty} (A_n-n\omega_b B_n)\\
    &\times\left\{\frac{ \cos(m t+n\omega_b\tilde t) }{m+n\omega_b} +\frac{ \cos(m t-n\omega_b\tilde t) }{m-n\omega_b}\right\}
    +H^{\Theta'}_{P_b,Y}(t)\,,\\
    H^{\Theta'}_{P_b,X}(t)=&-\frac{ \Phi_0  \omega_b}{2^{3/2}}\left\{\frac{-7  \sin m t }{m}  +   \sum_{n=1}^{+\infty}\frac{6 J_n(n e)}{(n \omega_b-m) (m+ n\omega_b)}\right.\nonumber\\
    &\times \left[ \frac{(m-n \omega_b) (2 m- n\omega_b) \sin (m t+t n \omega_b-T_0  n\omega_b )}{m+ n\omega_b}\right. \nonumber\\
    &+ \left.\left.  \frac{(m+ n\omega_b) (2 m+ n\omega_b) \sin (m t-t n \omega_b+T_0  n\omega_b )}{m-n \omega_b}\right]\right\} \,,\\
    H^{\Theta'}_{P_b,Y}(t)=&-\frac{  \Phi_0  \omega_b}{2^{3/2}}\left\{\frac{-7 \cos m t }{m}  +   \sum_{n=1}^{+\infty}\frac{6 J_n(n e)}{(n \omega_b-m) (m+ n\omega_b)}\right.\nonumber\\
    &\times \left[ \frac{(m-n \omega_b) (2 m- n\omega_b) \cos (m t+t n \omega_b-T_0  n\omega_b )}{m+ n\omega_b}\right. \nonumber\\
    &+ \left.\left.  \frac{(m+ n\omega_b) (2 m+ n\omega_b) \cos (m t-t n \omega_b+T_0  n\omega_b )}{m-n \omega_b}\right]\right\} \,,   \\  
    H^a_X(t)=& -\sqrt{2} \Phi_0 a  \cos m t-4\sqrt{2}    \Phi_0  a\sum_{n=1}^{+\infty} J_n(ne) \cos m t \cos(n\omega_b\tilde t)\nonumber\\
    &+3\sqrt{2}    \Phi_0 a \omega_b \sum_{n=1}^{+\infty} n J_n(ne)\left\{\frac{\cos(m t+n\omega_b\tilde t)}{m+n\omega_b } -  \frac{\cos(m t-n\omega_b\tilde t)}{m-n\omega_b } \right\}\,,\\
    H^a_Y(t) =& \sqrt{2}  \Phi_0  a \sin m t  +4\sqrt{2}    \Phi_0  a\sum_{n=1}^{+\infty} J_n(ne)  \sin m t\cos(n\omega_b\tilde t) \nonumber\\
    &-3\sqrt{2}     \Phi_0  a\omega_b \sum_{n=1}^{+\infty} n J_n(ne)\left\{\frac{\sin(m t +n\omega_b\tilde t)}{m+n\omega_b } -  \frac{\sin(m t-n\omega_b\tilde t)}{m-n\omega_b } \right\}\,,     \\  
    H^e_X(t)=& \Phi_0  \frac{1-e^2}{\sqrt{2}e}\sum_{n=1}^{+\infty} J_n(ne) \left\{ \frac{n\omega_b-2m}{ n\omega_b+m} \cos(m t+n\omega_b \tilde t) +\frac{2m+n\omega_b}{n\omega_b-m}\cos(m t-n\omega_b \tilde t) \right\}\, ,\\
    H^e_Y(t)=& -\Phi_0  \frac{1-e^2}{\sqrt{2} e}\sum_{n=1}^{+\infty} J_n(ne) \left\{ \frac{n\omega_b-2m}{ n\omega_b+m}\sin(m t+n\omega_b \tilde t)  +\frac{2m+n\omega_b}{n\omega_b-m} \sin(m t -n\omega_b \tilde t) \right\}\,, \\  
    H^\omega_X(t)=&  \Phi_0 \frac{\sqrt{1-e^2}}{\sqrt{2} e} \sum_{n=1}^{+\infty}J_n'(ne) \left\{ \frac{n\omega_b-2m}{ n\omega_b+m} \sin(m t+n\omega_b \tilde t)  -\frac{2m+n\omega_b}{n\omega_b-m} \sin(m t-n\omega_b \tilde t) \right\}\,,\\
    H^\omega_Y(t)=&  \Phi_0 \frac{\sqrt{1-e^2}}{\sqrt{2} e} \sum_{n=1}^{+\infty}J_n'(ne) \left\{ \frac{n\omega_b-2m}{ n\omega_b+m} \cos(m t+n\omega_b \tilde t)   -\frac{2m+n\omega_b}{n\omega_b-m} \cos(m t-n\omega_b \tilde t) \right\}\,;
\end{align}
\end{subequations}
}
for $a$, $e$ and $\omega$, the $\ell^S(t)$ contributions vanish while  $\ell^{\Theta'}(t)$ involves a linearly growing  term
\begin{equation}\label{deltaTheta}
 L^{\Theta'}_X(t) = \frac{3\omega_b}{2 a}    H_X^a(T_0)  \tilde t\,,\,\,\,\,\,\,L^{\Theta'}_Y(t) = \frac{3\omega_b}{2 a}  H_Y^a(T_0)   \tilde t. 
\end{equation}
As already mentioned, in the two-step approach, in the definition of $h^S(t)$ a constant and a term that is linearly growing (in time) are irrelevant since they can be absorbed into a redefinition of the nuisance parameters. The reason we keep the constant terms in $h^S(t)$ (given by $H_{X}^S(T_\text{ref})$ and $H_{X}^S(T_\text{ref})$) is because some variations present resonances and the constant parts are needed to obtain finite results at the resonance points.

\paragraph*{Perturbed ELL1 model} The model for the variations of the orbital parameters for near-circular orbits including the ULDM contribution is given by:
{\allowdisplaybreaks
\begin{subequations}\label{signals}
\begin{align}
    \delta\Psi'(t) =& \delta\Psi'_\text{asc}-\frac{3\omega_b}{2}\left(\frac{\delta a_\text{asc}}{a}+2  \alpha\Phi_0 \varrho\cos(m T_\text{asc}+\Upsilon) \right)\bar t-\frac{3\omega_b}{4}\frac{\dot{a}_\text{asc} }{a}\bar t^2 \nonumber\\
    &+\frac{11\omega_b}{2 m}\alpha \Phi_0 \varrho \left[\sin(m t+\Upsilon)-\sin(m T_\text{asc}+\Upsilon)\right] \,,\label{deltaPsiprime} \\
    \delta{x}(t) =& \delta{x}_\text{asc}+\dot{x}_\text{asc} \bar t-2 \alpha x \Phi_0 \varrho [\cos(m t+\Upsilon)-\cos(m T_\text{asc}+\Upsilon)] \,, \label{deltaX}\\
    \delta{\kappa}(t) =& \delta{\kappa}_\text{asc}+ \dot{\kappa}_\text{asc} \bar t + \alpha   \Phi_0 \varrho \frac{\kappa }{\kappa^2+\eta^2}  \left\{  \frac{\kappa (\omega_b+2 m)}{2(\omega_b-m)}\left[ \cos(m t-\omega_b\bar t + \Upsilon) -\cos(m T_\text{asc} + \Upsilon) \right] \right.\,  \nonumber\\
    &+\left. \frac{\kappa (\omega_b-2 m)}{2(\omega_b+m)}\left[ \cos(m t+\omega_b\bar t + \Upsilon) -\cos(m T_\text{asc} + \Upsilon) \right]   -   \frac{\eta(\omega_b+2 m)}{2(\omega_b-m)}\left[ \sin(m t-\omega_b\bar t + \Upsilon) \right.\right.\nonumber\\
    &-\left.\left.\sin(m T_\text{asc} + \Upsilon) \right]+  \frac{\eta(\omega_b-2 m)}{2(\omega_b+m)}\left[ \sin(m t+\omega_b\bar t + \Upsilon) -\sin(m T_\text{asc} + \Upsilon) \right]  \right\}  \,, \label{deltaKappa}\\
    \delta{\eta}(t) =& \delta{\eta}_\text{asc}+ \dot{\eta}_\text{asc} \bar t + \alpha   \Phi_0 \varrho \frac{\eta }{\kappa^2+\eta^2}  \left\{  \frac{\kappa (\omega_b+2 m)}{2(\omega_b-m)}\left[ \cos(m t-\omega_b\bar t + \Upsilon) -\cos(m T_\text{asc} + \Upsilon) \right] \right.\,  \nonumber\\
    &+\left. \frac{\kappa (\omega_b-2 m)}{2(\omega_b+m)}\left[ \cos(m t+\omega_b\bar t + \Upsilon) -\cos(m T_\text{asc} + \Upsilon) \right]   -   \frac{\eta(\omega_b+2 m)}{2(\omega_b-m)}\left[ \sin(m t-\omega_b\bar t + \Upsilon) \right.\right.\nonumber\\
    &-\left.\left.\sin(m T_\text{asc} + \Upsilon) \right]+  \frac{\eta(\omega_b-2 m)}{2(\omega_b+m)}\left[ \sin(m t+\omega_b\bar t + \Upsilon) -\sin(m T_\text{asc} + \Upsilon) \right]  \right\}  \,, \label{deltaEta}
\end{align}
\end{subequations}
}
where $\bar t=t-T_\text{asc}$, the quantities \(\delta a_\text{asc}\), \(\delta x_\text{asc}\), \(\delta \kappa_\text{asc}\) and \(\delta \eta_\text{asc}\) account for possible errors evaluated at \(T_\text{asc}\) and, as for the BT model, possible secular variations due to other effects are also included in the model. The corresponding $X$ and $Y$ components defined in \autoref{hsignal}, where $T_\text{ref}=T_\text{asc}$ in the ELL1 model, are given by
{\allowdisplaybreaks
\begin{subequations} 
\label{PerturbELL1}
\begin{align}
    H^{\Psi'}_X(t)=& \frac{11\omega_b}{2^{3/2} m}   \Phi_0  \sin (m t)\,,
    \,\,\,\,\,\,\,\,\,\,
    H^{\Psi'}_Y(t)=   \frac{11\omega_b}{2^{3/2} m}   \Phi_0 \cos (m t) \,,\label{Psiprimesignal}\\ 
    H^{x}_X(t)=&- \sqrt{2} x  \Phi_0   \cos (m t) \,,\,\,\,\,\,\,\,\,\,\,
    H^{x}_Y(t)=\sqrt{2} x  \Phi_0   \sin (m t) \,, \\
    H^\kappa_X(t)=&   \frac{  \Phi_0   \kappa }{2^{3/2}(\kappa^2+\eta^2)}  \left\{  \frac{ (\omega_b+2 m)}{(\omega_b-m)}\left[   \kappa  \cos(m t-\omega_b\bar t )    - \eta \sin(m t-\omega_b\bar t ) \right] \right. \nonumber\\
    &+ \left.  \frac{ (\omega_b-2 m)}{(\omega_b+m)}\left[  \kappa   \cos(m t+\omega_b\bar t )  +  \eta   \sin(m t+\omega_b\bar t )   \right] \right\} \,,\\
    H^\kappa_Y(t)=&      \frac{   \Phi_0   \kappa }{2^{3/2}(\kappa^2+\eta^2)}  \left\{  \frac{- (\omega_b+2 m)}{(\omega_b-m)}\left[  \eta \cos(m t-\omega_b\bar t )  +\kappa   \sin(m t-\omega_b\bar t ) \right] \right.
    \nonumber\\
    &+\left.\frac{ (\omega_b-2 m)}{(\omega_b+m)}\left[   \eta \cos(m t+\omega_b\bar t )    -   \kappa\sin(m t+\omega_b\bar t )  \right] \right\} \,,\\
    H^\eta_X(t)=&\frac{\eta}{\kappa} H^\kappa_X(t)\,,\,\,\,\,\,\,\,\,\,\,H^\eta_Y(t)=\frac{\eta}{\kappa} H^\kappa_Y(t)\,.
\end{align}
\end{subequations}
}
The linear terms for $x,\,\kappa$ and $\eta$, are zero, $\ell^S=0$, while for $\Psi'$ we have
\begin{equation}\label{deltaPsiL}
    L^{\Psi'}_X(t)= -\frac{3\omega_b}{2^{1/2}}   \Phi_0 \cos(m T_\text{asc})  \bar t\,,\,\,\,\,\,\, L^{\Psi'}_Y(t) = \frac{3\omega_b}{2^{1/2}}   \Phi_0 \sin(m T_\text{asc}) \bar t. 
\end{equation}

Notice that the signals for the parameters \(\eta\) and \(\kappa\) exhibit a resonant term, which comes from the first resonance \(m=\omega_b\) and is the only one that survives in the \(e\rightarrow0\) limit.  

\paragraph*{Integrals for marginalisation}
The integral expression for the quantities needed to evaluate the Bayes' factor $\mathcal{B}$ are:
\begin{subequations} 
\label{expforus}
\begin{align}
u&=\int_0^{T_{obs}}\frac{  (G h^S)^2(t)}{2 \sigma_{S}^2 d}  dt\,,\\
u_X&=\int_0^{T_{obs}}\frac{  (G A_{X}^S)(t) (G h^S)(t)}{2 \sigma_{S}^2 d}  dt\,,\\
u_Y&=\int_0^{T_{obs}}\frac{ (G A_{Y}^S)(t) (G h^S)(t)}{2 \sigma_{S}^2 d} dt\,,
 \end{align}
\end{subequations}
where all quantities are fixed to their fiducial values and  for a given function of time $f(t)$,
\begin{eqnarray}
    (Gf)(t)&=&f(t)-\sum_{i=m_S}^{2} f_i(t)\frac{\int_{0}^{T_{obs}} f_{i}(\tau) f(\tau)\,d\tau}{\int_{0}^{T_{obs}} f_{i}^2(\tau)  \,d\tau}\,, \nonumber\\
    f_{0}(\tau)&=&\frac{\tau^2}{T_{obs}}-\tau+\frac{T_{obs}}{6},\,\,\,\,f_{1}(\tau)= \tau-\frac{T_{obs}}{2}\,\,\,\,f_{2}(\tau)= T_\text{obs} \,,
\end{eqnarray}
where $m_S=0$ if $\Xi_2\neq 0$, $m_S=1$ if $\Xi_2=0$ and $\Xi_1\neq0$, and $m_S=2$ if $\Xi_2=0=\Xi_1$ and only $\Xi_0\neq0$.

%-------------------------------------------------------------------------------
\section{Resonances: going beyond $P_b$ in the BT model}\label{app:beyondPb}

In this appendix we discuss on  the importance of taking into account  the ULDM effect on orbital parameters other than the orbital period $P_b$ (which is the most studied one) for eccentric systems. We focus here on   system J1903+0327, whose parameters are given in \autoref{tab:3_high_e_PSRs}, and compute the sensitivity curves using the effect of ULDM on the orbital parameter $\Theta'$. The variance of $\delta\Theta'$ is given in \autoref{VarThetaprime} while the model  can be computed from \autoref{eq:varThpAp} where $A_n$ and $B_n$ are defined in \autoref{AnBn}. 

Upon using \autoref{eq:vardeltaaAp} into \autoref{eq:varThpAp} we can write $\delta\Theta'$ as
\begin{equation}\label{modelThpAp}
    \delta\Theta'=\Xi_0+\Xi_1 t-2 \pi \frac{\dot{P_b}_0}{P_b^2}t^2+h^{\Theta'}(t)\,, 
\end{equation}
where we have used the fact that
\begin{equation}\label{apb}
    \frac{\dot{a}}{a}=\frac{2}{3}\frac{\dot{P_b}}{P_b}+\frac{\alpha}{3}\dot{\Phi}\,,
\end{equation}  and we have included the parameter $\dot{P_b}_0$ in order to describe all possible secular variations that are not caused by ULDM (e.g.\ the Shklovskii effect~\cite{1970SvA....13..562S} or the galactic acceleration~\cite{1991ApJ...366..501D}). Here we collect in $\Xi_0$ all the constant contributions to $\delta\Theta'$, regardless of whether these come from ULDM or some other effects; all the contributions growing linearly with $t$ are included in the $\Xi_1 t$ term.  The relevant ULDM signal is characterised by the functions $h^{\Theta'}(t)$ and $h^{\Theta'}_{P_b}(t)$ given in
\autoref{app:bayes} as
\begin{subequations}\label{hsignalTheta}
\begin{align}\label{hsignalThetafull}
h^{\Theta'}(t)&= \alpha [( H_{X}^{\Theta'}(t)- H_{X}^{\Theta'}(T_0))X+( H_{Y}^{\Theta'}(t)- H_{Y}^{\Theta'}(T_0)) Y]\,,\\ 
h^{\Theta'}_{P_b}(t)&= \alpha [( H_{P_b,X}^{\Theta'}(t)- H_{P_b,X}^{\Theta'}(T_0))X+( H_{P_b,Y}^{\Theta'}(t)- H_{P_b,Y}^{\Theta'}(T_0)) Y]\,,
\end{align}
\end{subequations} where  the functions on the right-hand-side are\label{htheta} defined in \autoref{hthetaXY}.
Here  $h^{\Theta'}(t)$ describes  the effect of ULDM on $\Theta'$ given in \autoref{eq:varThpAp} and  $h^{\Theta'}_{P_b}(t)$   takes into account only the ULDM effect on $P_b$   obtained from the last term in \autoref{eq:varThpAp}  and \autoref{apb}.

Notice that the constant parameters $\Xi_0$  includes the error in the determination of $\Theta'_0$, and the coefficient $\Xi_1$ of the linear term  depends on the error in the determination of $a_0$. In order to obtain the sensitivity curve for $\alpha$, we use the variance of $\delta\Theta'$ given in \autoref{VarThetaprime} and marginalise over the two nuisance parameters $\Xi_0$ and $\Xi_1$ in the model of $\Theta’$ using uniform priors for both of them. We assume here that any possible secular effect on $P_b$ characterised by $\dot{P_b}_0$ in \autoref{modelThpAp} can be independently determined with significant precision so that we can subtract them and then set $\dot{P_b}_0=0$ (see \autoref{AppShape} where we show that this simplification is only relevant for masses that are very close to the resonance; this is the case because ULDM produces a secular effect at $m=n\omega_b$, but not for other values of the mass).

\begin{figure}
    \centering
    \includegraphics[width=\textwidth]{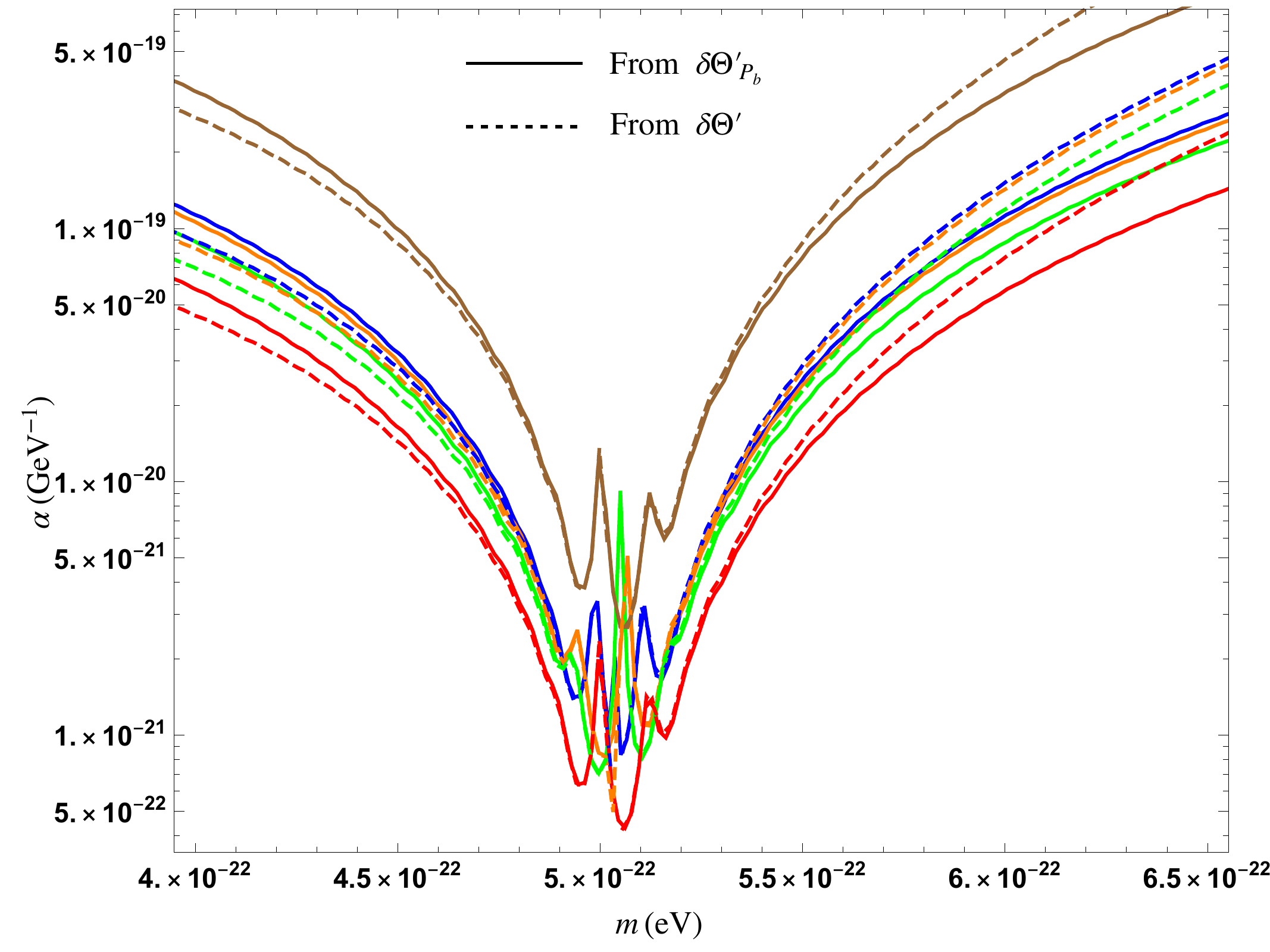}
    \caption{Sensitivity curves for $\alpha$ as a function of the ULDM mass $m$ for the system J1903+0327 (with parameters given in \autoref{tab:3_high_e_PSRs}) for masses near the resonant one at $\omega_b=m$. Dashed lines account for all the ULDM effects on the parameter $\Theta'$, while the solid lines are obtained neglecting the ULDM effect on all parameters but $P_b$. Each colour corresponds to a particular random realisation of the parameters $\varrho$ and $\Upsilon$.}\label{sensitivRes4RealThetavsPv}
\end{figure}

In order to illustrate the difference in sensitivity that could be obtained by performing an analysis with $h^{\Theta'}(t)$ in comparison to   $h^{\Theta'}_{P_b}(t)$,  we restrict ourselves to the study of the signal at low masses, where the contribution of the first harmonic with $n=1$ in \autoref{eq:varThpAp} dominates the  signal and accounts for a resonant effect for $m=\omega_b$. Within the scope of this work it is enough to focus on the contribution of the first $n=1$ harmonic, so we neglect higher ones. The result is shown in \autoref{sensitivRes4RealThetavsPv}, where we plotted the sensitivity curves for five different random realisations of the values of $\varrho$ and $\Upsilon$, for which we used different colours. The dashed lines were obtained using $h^{\Theta'}(t)$ while the solid lines correspond to $h^{\Theta'}_{P_b}(t)$. We see the effect on $P_b$ dominates, so the estimated sensitivity obtained from both $h^{\Theta'}_{P_b}(t)$ and $h^{\Theta'}(t)$ are about the same. The use of $h^{\Theta'}_{P_b}(t)$ instead of  $h^{\Theta'}(t)$ underestimate  (overestimate) the sensitivity at the lower (higher) masses away from the resonant band but only by an order one factor.  

%-------------------------------------------------------------------------------
\section{Marginalisation and the resonant peaks in the two-step approach}\label{AppShape}

In this appendix we analyse the sensitivity curves near the resonances. To this end, we consider the effect on $\delta\Theta'$ at low masses around the first resonant peak $m=\omega_b$ and neglect higher harmonics. For the sake of generality, we model the signal for the second step as
\begin{equation}
    \delta\Theta'(t)= \Xi_0+\Xi_1 t+\Xi_2 t^2+ h^{\Theta'}(t)+\ell^{\Theta'}(t) \,,
\end{equation}
where $h^{\Theta'}(t)$ and $\ell^{\Theta'}(t)$ are defined as in \autoref{hsignal} and are given in \autoref{hsignalThetafull} and \autoref{deltaTheta}, respectively. Here the parameters $\Xi_i$  ($i=0,1,2$) do not include the linear function in time coming from the ULDM contribution, which is given by $\ell^{\Theta'}(t)$, nor the secular quadratic term that shows up at the resonant mass.

\begin{figure}
    \centering
    \includegraphics[width=\textwidth]{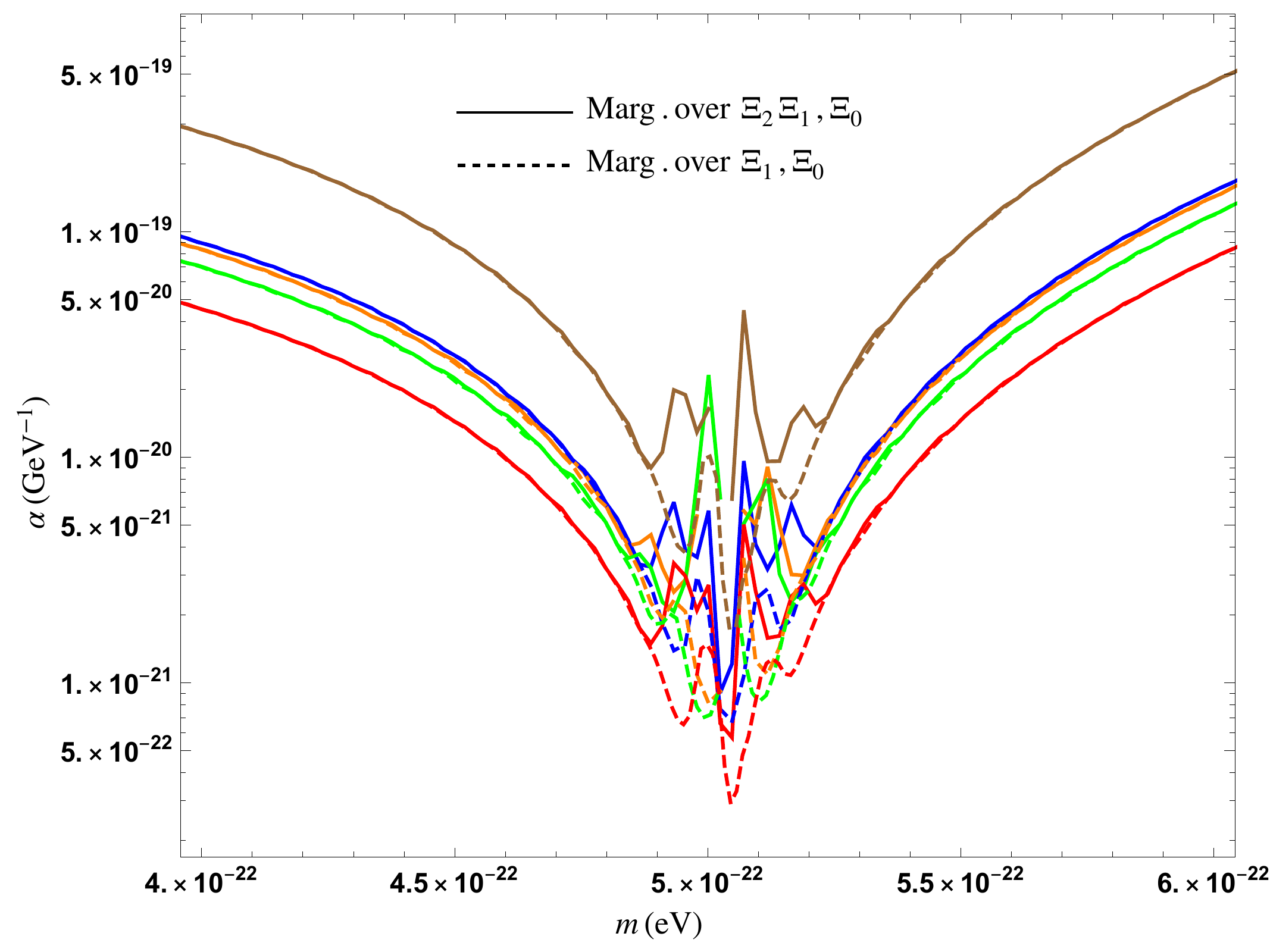}
    \caption{Sensitivity curves for $\alpha$ as a function of the ULDM mass $m$ for the system J1903+0327 (with parameters given in \autoref{tab:3_high_e_PSRs}) for masses near the resonant one  at $\omega_b=m$. Solid lines are obtained after marginalising over all the three nuisance parameters $\Xi_i$ ($i=0,1,2$) in the model of $\Theta’$ (we use uniform priors for all of them), dashed  lines are obtained after marginalising only over the parameters $\Xi_0$ and $\Xi_1$ (which correspond to a linear growth coefficient and a constant error in $\Theta’$) while the coefficient $\Xi_2$ (which gives a quadratic growth) is assumed to be known and subtracted so that $\Xi_2=0$.}\label{sensitivRes4Real}
\end{figure}

\autoref{sensitivRes4Real} shows the sensitivity curves as a function of the mass for the system J1903+0327 around the resonant mass $m=\omega_b$ for five realisations of the parameters $\varrho$ and $\Upsilon$. Firstly, we see that the shape of the curve around the resonant mass depends strongly on the values of the parameters $\varrho$ and $\Upsilon$. Secondly, since we do not know the  nuisance parameters $\Xi_i$  ($i=0,1,2$),  an appropriate procedure to obtain a conservative sensitivity curve for $\alpha$ is to marginalise over such irrelevant parameters. However, in \autoref{sensitivRes4Real} it can be seen that the differences between marginalising over the nuisance parameters, or neglecting them, are significant only in a tiny mass range near resonance.\footnote{Neglecting the nuisance parameters is equivalent to assuming that the parameters can be separately determined and then removed from the signal.} This is as expected, because only close to the resonance the ULDM produces a secular drift of the orbital period which can be swept in the marginalisation of the quadratic term $\Xi_2$.

\begin{landscape}
\vspace*{\fill}
\begin{table}[htbp]
\centering

\begin{tabular}{lrlrllrrrr}
\toprule
       PSR &   $P_b$ [d] &    $a_1$ [ls] &  $\omega$ [$^\circ$]&      $\eta$ [1] &      $\kappa$ [1] &  $T_{\mathrm{asc}}$ [MJD] &  $\epsilon$ [$\mu$s] &  $T_{\mathrm{obs}}$ [yr] &  $n_c$ [$\mathrm{yr}^{-1}$]\\
\midrule
J1745+1017 &   0.7 & 8.82e-02 &   212.5 & -8.95e-06 & -1.40e-05 & 58220.8 & 14.18 &   4.5 &     663.6 \\
J2145-0750 &   6.8 & 1.02e+01 &   201.0 & -6.91e-06 & -1.80e-05 & 56102.4 &  7.86 &  15.5 &    1201.9 \\
J1012+5307 &   0.6 & 5.82e-01 &    67.7 &  1.18e-06 &  4.87e-07 & 56104.0 &  7.94 &  15.5 &    1663.6 \\
J2317+1439 &   2.5 & 2.31e+00 &    93.3 &  4.73e-07 & -2.70e-08 & 56213.7 &  5.25 &  15.6 &     890.2 \\
J0557+1551 &   4.8 & 4.05e+00 &    96.0 &  8.59e-06 & -9.00e-07 & 58224.9 &  4.75 &   4.6 &     115.0 \\
J0509+0856 &   4.9 & 2.46e+00 &    31.8 &  1.16e-05 &  1.88e-05 & 58403.8 & 12.36 &   3.6 &     608.7 \\
J1802-2124 &   0.7 & 3.72e+00 &    51.7 &  2.65e-06 &  2.10e-06 & 58314.1 &  8.14 &   3.5 &    1965.9 \\
  B1855+09 &  12.3 & 9.23e+00 &   276.5 & -2.16e-05 &  2.44e-06 & 56211.1 &  3.29 &  15.6 &     497.8 \\
J2017+0603 &   2.2 & 2.19e+00 &   179.8 &  2.14e-08 & -6.99e-06 & 57510.9 &  2.68 &   8.3 &     421.9 \\
J1614-2230 &   8.7 & 1.13e+01 &   176.1 &  8.96e-08 & -1.33e-06 & 56830.8 &  3.23 &  11.5 &    1598.2 \\
J2234+0944 &   0.4 & 6.84e-02 &     4.9 &  5.72e-07 &  6.62e-06 & 57755.1 &  5.76 &   7.1 &    1061.4 \\
J2043+1711 &   1.5 & 1.62e+00 &   237.9 & -4.26e-06 & -2.67e-06 & 57413.5 &  1.59 &   9.1 &     817.2 \\
J1125+7819 &  15.4 & 1.22e+01 &   274.0 & -1.28e-05 &  8.89e-07 & 57785.2 & 13.85 &   6.3 &    1385.5 \\
J1811-2405 &   6.3 & 5.71e+00 &    62.9 &  1.04e-06 &  5.34e-07 & 58316.2 &  3.12 &   3.5 &    1523.3 \\
J0740+6620 &   4.8 & 3.98e+00 &   252.0 & -5.65e-06 & -1.84e-06 & 57795.2 &  4.44 &   6.3 &    2123.5 \\
J1719-1438 &   0.1 & 1.82e-03 &    66.3 &  3.63e-04 &  1.59e-04 & 58312.0 & 15.80 &   3.4 &    1846.1 \\
J0406+3039 &   7.0 & 2.32e+00 &   161.4 &  3.95e-06 & -1.17e-05 & 58404.8 &  2.98 &   3.6 &     686.5 \\
J0613-0200 &   1.2 & 1.09e+00 &    32.9 &  2.28e-06 &  3.52e-06 & 56196.4 &  3.26 &  15.0 &    1138.4 \\
J1741+1351 &  16.3 & 1.10e+01 &   204.0 & -4.06e-06 & -9.12e-06 & 57060.7 &  2.40 &  11.0 &     506.8 \\
J1738+0333 &   0.4 & 3.43e-01 &   107.0 &  1.07e-06 & -3.27e-07 & 57096.2 &  5.06 &  10.7 &     819.0 \\
J1909-3744 &   1.5 & 1.90e+00 &   155.8 &  4.46e-08 & -9.91e-08 & 56121.0 &  1.18 &  15.5 &    2261.6 \\
J0610-2100 &   0.3 & 7.35e-02 &   120.9 &  1.32e-05 & -7.92e-06 & 58328.1 &  9.41 &   3.4 &    1451.1 \\
\bottomrule
\end{tabular}

\caption{22 pulsars described by the ELL1(H) model.}
\label{tab:22_ELL1psrs}

\end{table}
\vspace*{\fill}
\end{landscape}
\begin{table}[htbp]
\centering

\begin{tabular}{lrrrrrrrr}
\toprule
       PSR &  $P_b$ [d] &  $a_1$ [ls] &  $\omega$ [$^\circ$] &   $e$ [1] &    $T_0$ [MJD] &  $\epsilon$ [$\mu$s] &  $T_{\mathrm{obs}}$ [yr] &  $n_c$ [$\mathrm{yr}^{-1}$] \\
\midrule
J1946+3417 &  27.0 &   13.9 &   223.4 & 0.134 & 58008.3 &  3.47 &   5.7 &     827.2 \\
J2234+0611 &  32.0 &   13.9 &   277.2 & 0.129 & 57850.1 &  3.25 &   6.5 &     545.5 \\
J1903+0327 &  95.2 &  105.6 &   141.7 & 0.437 & 57109.4 &  6.24 &  10.7 &     638.8 \\
\bottomrule
\end{tabular}

\caption{Three pulsars with notably high eccentricity parametrised by the BT timing model.}
\label{tab:3_high_e_PSRs}

\end{table}

%-------------------------------------------------------------------------------
\section{A one-step approach to estimate the sensitivity (cross-verification)}\label{app:xcz}

In this section we fit time residuals directly to the timing model (\textit{one-step approach}) in order to construct the sensitivity curves. This approach is based on less robust assumptions, in particular we assume here that time residuals are due \textit{solely} to the effect of ULDM on the binary's orbital parameters and are subject to a Gaussian (white) noise distribution. We focus our analysis on the 22 ELL1(H) pulsars included in the NANOGrav 2023 dataset (see \autoref{tab:22_ELL1psrs}) and then extend this analysis to the three binary pulsars with higher eccentricity of \autoref{tab:3_high_e_PSRs}.

\paragraph*{Disadvantages.} In the two-step approach we have introduced nuisance parameters to model possible errors or constant shifts in the determination of the orbital parameters and secular drifts from other sources, over which we have marginalised. The one-step method we describe here does not take into account any effect other than ULDM and noise, i.e.\ no such errors and secular drifts are included. Consequently, the sensitivity curves produced by this approach tend to be more optimistic (or less conservative) compared to those derived from the two-step method. 
 
\paragraph*{Advantages.} This method allows us to jointly consider variations of all orbital parameters, enhancing sensitivity---a capability lacking in the previous method. The marginalisation over $\varrho$ and $\Upsilon$ using a Gaussian prior can be done fully analytically, giving insight into how the sensitivity depends on each variable such as observation time, noise magnitude, or cadence. Importantly, plotting the sensitivity curves is not time consuming, compared to the two-step method. Lastly, since not all binary pulsars are of the same quality in terms of their ability to detect ULDM, this method can be used to quickly determine which pulsars are the most suitable to study ULDM, allowing us to restrict ourselves to a selected subset of pulsars. This knowledge can be further exploited when using the two-step method.

\subsubsection*{Timing model}
In line with our assumptions, we decompose time residual $R$ into two components
\begin{equation}
\label{eq:time_resid_DM_and_noise}
     R = R^{\mathrm{DM}} + R^{\mathrm{PN}} \,.
\end{equation}
Here, $R^{\mathrm{DM}}$ represents the contribution of ULDM in the ELL1 model, while $R^{\mathrm{PN}}$ accounts for pulsar noise. Within the ELL1 model given in \autoref{ResELL1}, at leading order in the small quantities $\eta$, $\kappa$, $x\omega_b$ and $t \dot{\nu}/\nu$, the variation of $\delta x, \delta \Psi', \delta \eta, \delta \kappa$ leads to the following expression for the time residual $R^{\mathrm{ELL1}}$
\begin{equation}\label{ELL1Pav}
    R^{\mathrm{ELL1}} = \sin \Psi' \delta x + x \cos \Psi' \delta \Psi' - x \frac{\cos(2\Psi')+3}{2} \delta \eta + x \frac{\sin(2 \Psi')}{2} \delta \kappa \,,
\end{equation}
while $R^{\mathrm{ELL1}}$ transforms into $R^{\mathrm{DM}}$ when $\delta x, \delta \Psi', \delta \eta, \delta \kappa$ arise from ULDM, given by  \autoref{signals}. Since we are assuming here that the variations only come from the ULDM, in such perturbation we set to zero both the error and the derivative of the orbital parameters at $T_{\text{asc}}$, i.e.\ $\delta x = h^x$ etc.

Notice that the function $\delta \Psi'(t)$ given in \autoref{deltaPsiprime} includes a constant term and a linear term (in time) that depend on the ULDM field, while the ULDM signal considered in the two-step method, namely $h^{\Psi'}(t)$ defined in \autoref{hsignal}, does not incorporate those contributions, since they have been absorbed into the nuisance parameters. The presence of the linear term significantly enhances the sensitivity, but the result should be taken with a grain of salt because it assumes that other linear contributions are either subdominant or can be independently measured and subtracted away, whereas we expect that it is indistinguishable from an error in the semi-major axis \(a_1\). Nonetheless, in order to understand the physical effect of the linear term, we will present the sensitivity plots using the one-step method both with the presence of the linear term and without it. The sensitivity curves obtained in this section (without the linear term) become very close to the ones obtained from the two-step method.

\subsubsection*{Noise model}
Since we have already described dark matter component (see, e.g.\ \autoref{app:bayes}), let us turn our attention to the noise model. We consider uncorrelated white Gaussian noise with correlation matrix
\begin{equation}
    C_{ai,bj}=\epsilon_a^2 \delta_{ab} \delta_{ij} \,,
\end{equation}
where \(a,b\) are pulsar indices and \(i,j\) are TOAs indices, $\epsilon_a$ is the noise amplitude of pulsar $a$ and the \(\delta\)s are Kronecker deltas. % $i$ is the time of arrival (TOA) of its pulse. 

Concentrating on a single pulsar $a$, the likelihood function $P_a$ of time residuals, denoted as $\{R_i\}_a$, where curly brackets denote a set in which $i = 1,\ldots,n_a$, with parameters $p_a\deq \left( \alpha, \varrho_a, \Upsilon_a \right)$ is
\begin{equation}
    P_a\left( \{R_{i}\}_a  | p_a \right)  = \frac{1}{ (2\pi)^{\frac{n_a}{2} }\epsilon_a^{n_a}} \exp \left[ - \frac{1}{2} \sum_{i=1}^{n_a} \frac{\left( R_{ai} - R_{ai}^{\mathrm{DM}} \right)^2}{\epsilon_a^2} \right] \,.
\end{equation}
The likelihood function combined for all pulsars $a=1,\dots,N_p$ is as follows
\begin{equation}
\label{eq:combined_prob_dist_func}
    P\left( \{R_{ai}\}  | p \right) \deq \Pi_{a=1}^{N_p} P_a\left( \{R_{i}\}_a | p_a \right) \,.
\end{equation}

\subsubsection*{Sensitivity limit ($\delta$-function priors)}
We apply Bayes' theorem to \autoref{eq:combined_prob_dist_func}
\begin{equation}
 P\left( p | \{R_{ai}\} \right) =  P\left( \{R_{ai}\}   | p \right) \frac{P_0 \left( p \right)}{P_0 \left( \{R_{ai}\} \right)} \,,
\end{equation}
where $P_0 \left( p \right)$ denotes the prior distribution for the unknown parameters $p \deq (\alpha, \{\varrho_a\}, \{\Upsilon_a\})$, and $P_0 \left( \{R_{ai}\} \right)$ is the evidence. In what follows, we consider $\delta$-function priors
\begin{equation}
    P_0 \left( p \right) = \Pi_{a=1}^{N_p} P_0 \left( p_a \right)  = \Pi_{a=1}^{N_p} \delta(\varrho_a - \varrho_a') \delta(\Upsilon_a - \Upsilon_a') \,,
\end{equation}
which leads to the probability distribution function of $\alpha$
\begin{equation}
    P\left( \alpha | \{R_{ai}\} \right) = \frac{1}{\sqrt{2 \pi} \sigma_{\alpha}} \exp\left[{-\frac{\left( \alpha  -  \alpha_{\mathrm{peak}}\right)^2}{2 \sigma_{\alpha}^2}} \right] \,,
\end{equation}
where the peak and standard deviation are
\begin{equation}
\label{eq:peak_and_sigma_1}
    \alpha_{\mathrm{peak}} \deq \sigma_{\alpha}^2\sum_{a=1}^{N_p}\frac{\alpha_{\mathrm{peak},a}}{\sigma_{\alpha,a}^2},~~\frac{1}{\sigma_{\alpha}^2} \deq \sum_{a=1}^{N_p} \frac{1}{\sigma_{\alpha,a}^2} \,,
\end{equation}
where
\begin{equation}
    \alpha_{\mathrm{peak},a} \deq \frac{\vec{R}_a . \vec{Q}_a}{\vec{Q}_a^2},~~\sigma_{\alpha,a} \deq \frac{\epsilon_a}{\sqrt{\vec{Q}_a^2}} \,,
\end{equation}
where $\vec{Q}_a\deq\vec{R}_a^{\text{DM}}/\alpha$ is a function describing the ULDM signal, but its magnitude is independent on the coupling constant (but still depends on $\varrho_a$ and $\Upsilon_a$). With this in hand, we are in a position to establish a simple sensitivity limit for the detection. Following  the well-known \textit{$3\sigma$-rule} for the Gaussian distribution, to ensure the exclusion of the null case the requirement 
\begin{equation}
\label{eq:six_sigma}
    |\alpha_f| >3 \sigma_{\alpha} \,,
\end{equation}
where $\alpha_f$ represents the genuine value in nature, should be met. Indeed, if we use the Bayes' factor to determine the detection limit, i.e.
\begin{equation}
    \mathcal{B} = \frac{\int P\left( \{R_{ai}\} | \alpha_f, \{\varrho_a\}, \{\Upsilon_a\} \right) P_0( \{\varrho_a\}, \{\Upsilon_a\}) \Pi_{a=1}^{N_p} d \varrho_a d \Upsilon_a}{P\left( \{R_{ai}\}| \alpha_{f}=0 \right)} \,,
\end{equation}
where $P\left( \{R_{ai}\}| \alpha_{f}=0 \right)$ is the likelihood function of pure white Gaussian noise (that is, the hypothesis that $\Vec{R}_a$ contains $\alpha_{f} = 0$), and we apply $\Vec{R}_a = \Vec{R}_a^{\mathrm{PN}} + \alpha_{f} \vec{Q}_a$\,, we find
\begin{equation}
\label{eq:Bayes_factor}
    \mathcal{B} \deq  \Pi_{a=1}^{N_p} \mathcal{B}_a \deq \Pi_{a=1}^{N_p}\exp\left[ \alpha_{f}^2 \frac{\Vec{Q}_a^2}{2 \epsilon_a^2} + \frac{\Vec{R}^{\mathrm{PN}}_a\cdot\Vec{Q}_a}{\epsilon_a^2}  \alpha_{f}  \right] \,.
\end{equation}
Notice that this is a noise-dependent quantity. To get rid of this dependency, we average over multiple noise realisations. This is not an unambiguous procedure; one way to average is as follows
\begin{equation}
\label{eq:alpha_true_delta}
    \bar{\mathcal {B}} \deq \int \mathcal {B} P\left( \{R_{ai}^{\mathrm{PN}}\} \right) \Pi_{a=1}^{N_p} d R_{a1}^{\mathrm{PN}} \dots d R_{an_a}^{\mathrm{PN}} = \exp\left[   \sum_{a=1}^{N_p}\frac{\Vec{Q}_a^2}{\epsilon_a^2}  \alpha_{f}^2 \right] \,,
\end{equation}
or we can average e.g.\ logarithms as
\begin{equation}
\label{eq:alpha_true_delta2}
    \ln \mathbb{B} \deq \int \ln \mathcal {B} P\left( \{R_{ai}^{\mathrm{PN}}\} \right) \Pi_{a=1}^{N_p} d R_{a1}^{\mathrm{PN}} \dots d R_{an_a}^{\mathrm{PN}} =  \sum_{a=1}^{N_p}\frac{\Vec{Q}_a^2}{2\epsilon_a^2}  \alpha_{f}^2 \,,
\end{equation}
yielding, in that order
\begin{align}
\label{eq:mathcal_delta}
    |\alpha_{f}| &= \sqrt{\ln \bar{\mathcal{B}}} \, \sigma_{\alpha} \doteq 2.63 \,\sigma_{\alpha} \,,\\
    |\alpha_{f}| &= \sqrt{2 \ln \mathbb{B}} \,\sigma_{\alpha} \doteq 3.72 \, \sigma_{\alpha} \,.
\end{align}
with $\bar{\mathcal{B}}=1000$ and $\mathbb{B} = 1000$. The take-home message is that, when we consider the $\delta$-function priors, the sensitivity limit is given by $\sigma_{\alpha}$, with a numerical factor close to 3. Throughout the following discussion, we will adhere to the factor of 3.

\begin{figure}[htb]
     \centering
     \includegraphics[width=13cm]{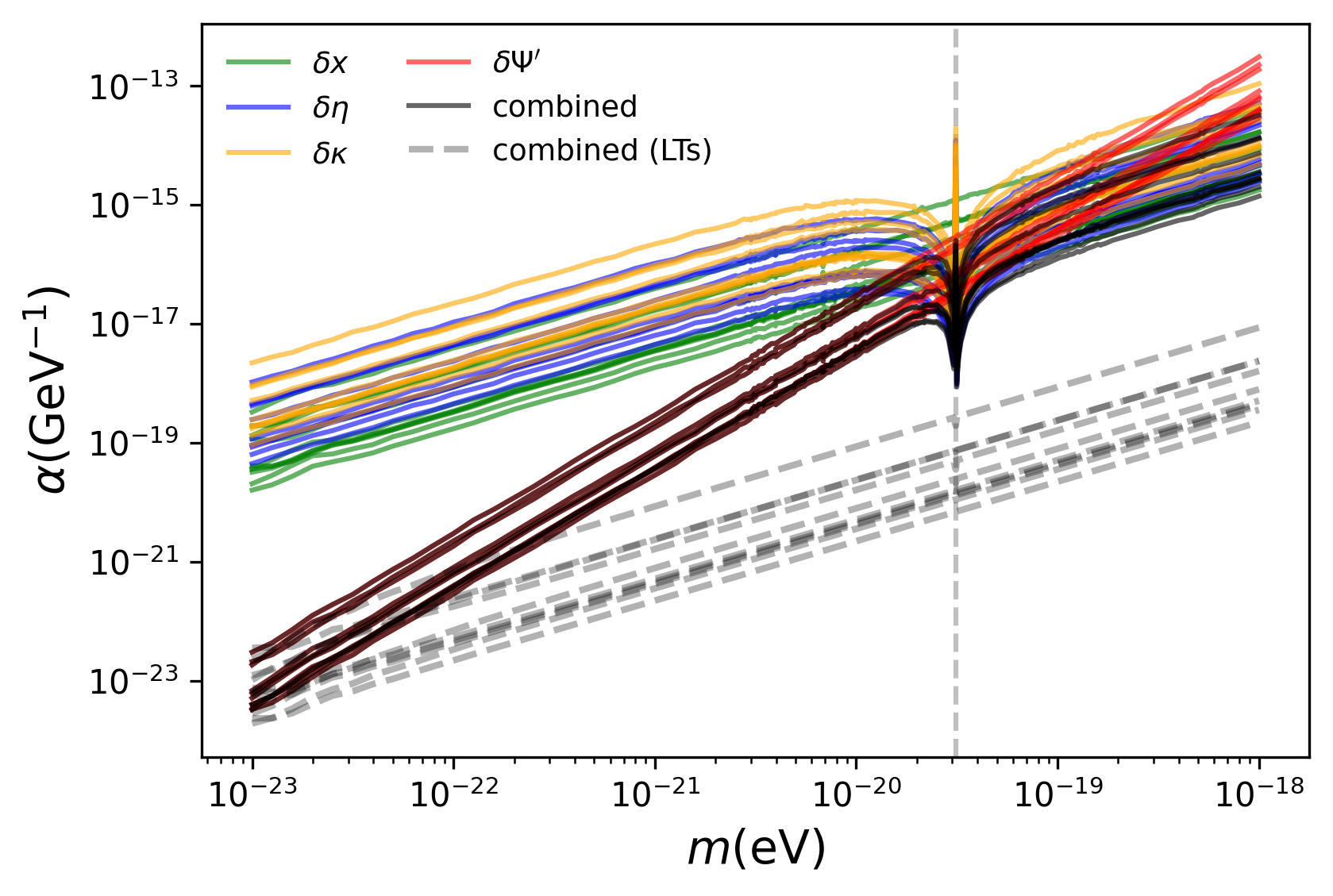}
     \caption{J1909-3744, sensitivity lines obtained for each parameter, ten random realisations. The vertical line corresponds to the orbital frequency of the binary. The black solid lines correspond to the combination of all parameters. Dashed lines correspond to the case when the linear terms (``LTs'') are kept.}
     \label{fig:1}
\end{figure}

\begin{figure}[htb]
     \centering
     \includegraphics[width=13cm]{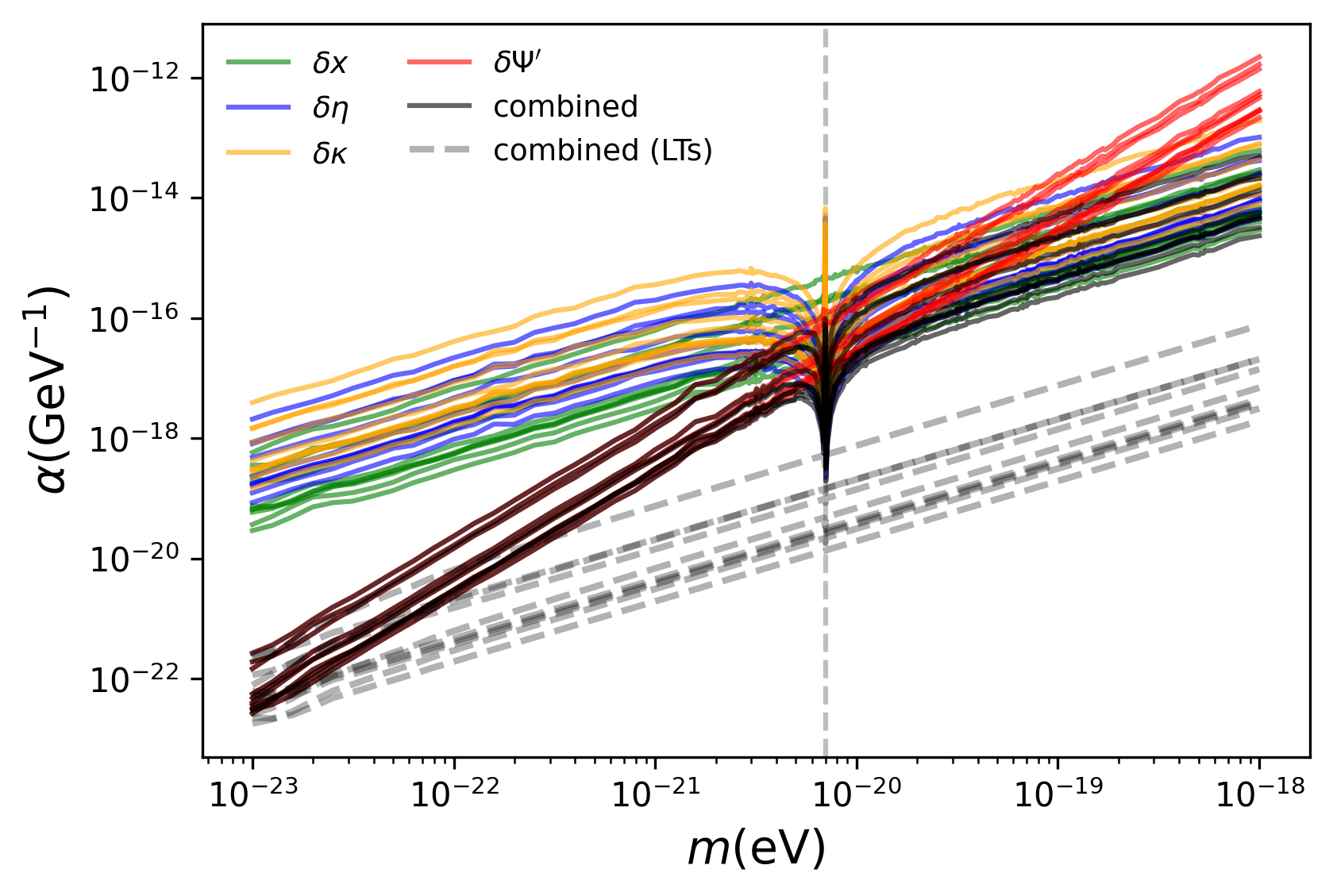}
     \caption{Same as \autoref{fig:1} but for the system J2145-0750.}
     \label{fig:2}
\end{figure}

\begin{figure}[htb]
     \centering
     \includegraphics[width=13cm]{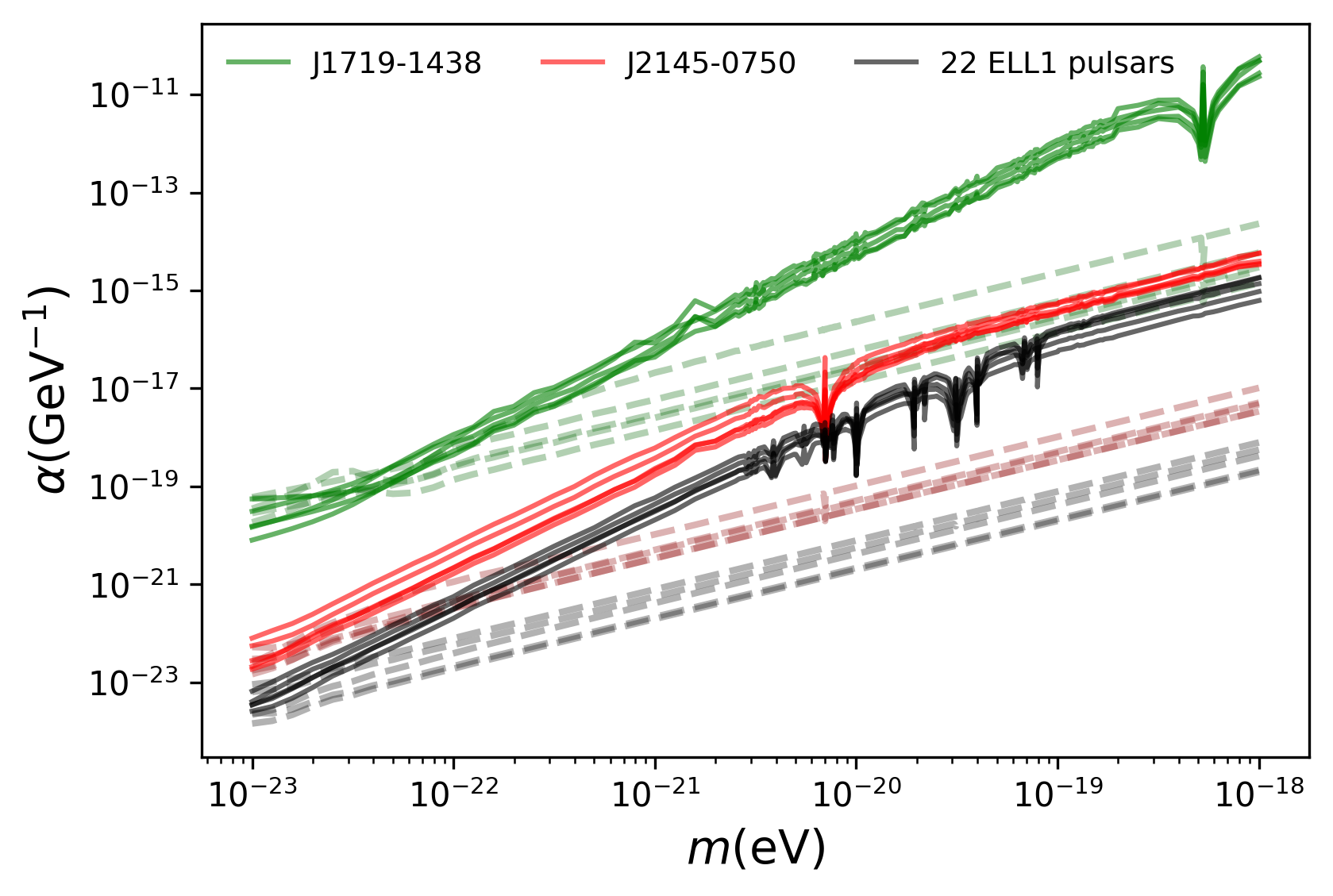}
     \caption{Five realisations of sensitivity lines using all orbital parameters. Solid lines correspond to cases where all linear terms have been removed, while dashed lines correspond to cases where they have been kept.}
     \label{fig:3}
\end{figure}

We show a few sensitivity lines for two selected pulsars in  \autoref{fig:1} and \autoref{fig:2}, and the combined result of all 22 ELL1(H) pulsars in \autoref{fig:3}. Indeed, we can observe that the so-called linear term has a significant effect on the estimated sensitivity derived from $\delta \Psi'$, both qualitatively (the shape of the curve) and quantitatively (the curve's normalisation). Fitting only the periodic component gives a less optimistic result and more similar to the results obtained by the two-step approach.

The parameters $\delta \eta$ and $\delta \kappa$ are of particular interest due to the resonance occurring when $m = \omega_b$. The depth of the signal relative to the line is approximately determined by the $\omega_b T$ factor, consistently around three to four orders of magnitude across all studied pulsars. This also suggests that the depth will scale with the length of the observation.

\subsubsection*{Large eccentricities}
For binary pulsars described by the BT model, the dominant part of the time residual (ignoring the contribution of $\delta K$) comes from \autoref{ResBT3}, whereas the variations due to ULDM can be obtained from \autoref{signalsBT} using the definition of $\alpha_b$ and $\eta_b$ in \autoref{defalphabetab} and recalling that $\eta_b\approx \beta_b$. Explicitly, 
\begin{align}
R^{\mathrm{BT}} =&\,
\frac{\delta x}{x}\left[\eta_b \sin  E'+\alpha_b\left(\cos E'-e\right)\right]- \delta e \left[\alpha_b +\frac{e \eta_b\sin E'}{1-e^2}+\frac{\sin E'  \left(\alpha_b \sin E'-\eta_b \cos E'\right)}{1-e \cos E'}\right]\nonumber\\
&+\delta \omega x\left[ -\sin E' \sin\omega (1-e^2)^{1/2} + \cos \omega \left(\cos E'-e\right)\right]-\frac{\delta \Theta ' \left(\alpha_b  \sin E'-\eta_b  \cos E'\right)}{1-e \cos E'} \,,
\end{align}where  the variations of $\delta x/x=\delta a/a$, $\delta\omega$, $\delta e$, and $\delta\Theta'$ are given by \autoref{signalsBT} and, as in the ELL1 case, in this appendix we  ignore  possible errors in the orbital parameters and secular drifts from other sources, and include  the linearly growing  term  in  $\delta \Theta '$  that depends on the ULDM field.

\begin{figure}[htb]
     \centering
     \includegraphics[width=13cm]{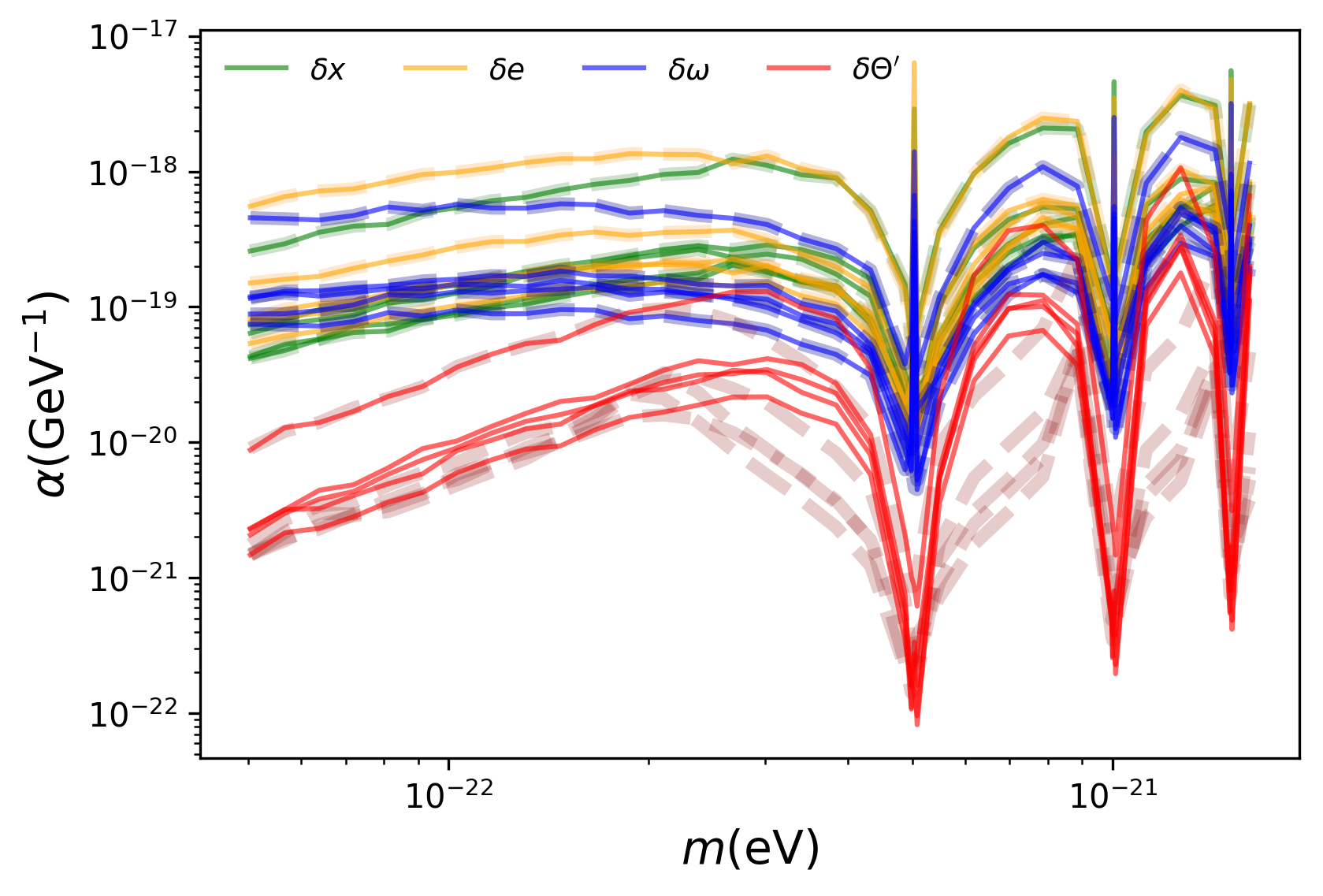}
     \caption{J1903+0327, five random realisations of sensitivity lines for parameters $\delta x$, $\delta e$, $\delta \omega$ and $\delta \Theta'$. Dashed lines correspond to the inclusion of linear terms.}
     \label{fig:more_eccentric_1}
\end{figure}

\begin{figure}[htb]
     \centering
     \includegraphics[width=13cm]{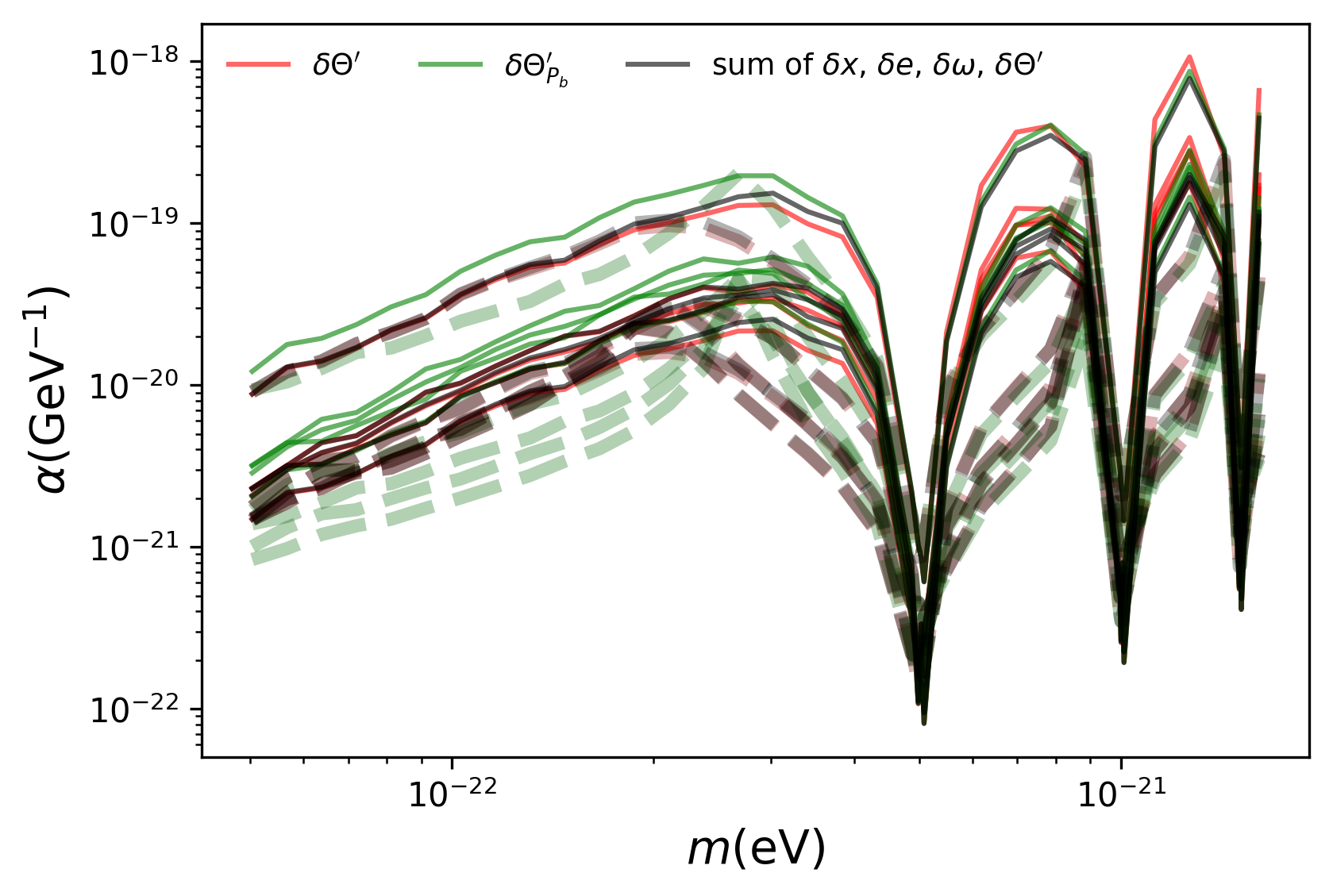}
     \caption{J1903+0327, five random realisations of sensitivity lines for $\delta \Theta'$, $\delta \Theta'_{P_b}$ and all combined parameters. Dashed lines correspond to the inclusion of linear terms.}
     \label{fig:more_eccentric_2}
\end{figure}

\begin{figure}[htb]
     \centering
     \includegraphics[width=13cm]{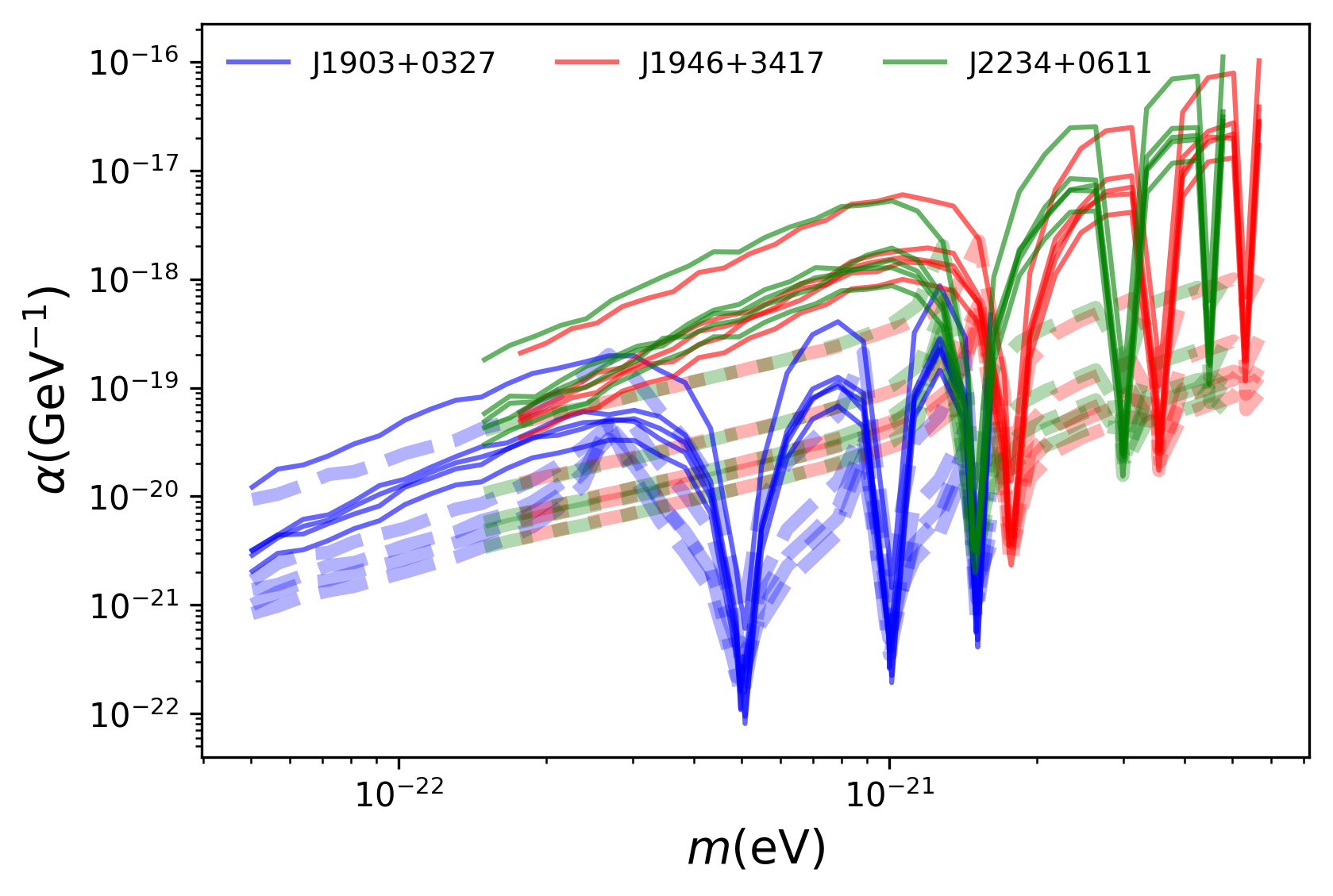}
     \caption{Five random realisations of sensitivity lines, derived from $\delta \Theta'_{P_b}$, for J1946+3417, J2234+0611 and J1903+0327. Dashed lines correspond to the inclusion of linear terms.}
     \label{fig:more_eccentric_3}
\end{figure}

Five sensitivity curves for $\delta x, \delta \omega, \delta e, \delta \Theta'$ are displayed in \autoref{fig:more_eccentric_1}. The shape of the sensitivity curves derived from $\delta \Theta'$ is non-trivial and can vary depending on the value of $\Upsilon$. The sensitivity obtained from the combination of all parameters, primarily influenced by $\delta \Theta'$, for all three high-eccentricity pulsars is depicted in \autoref{fig:more_eccentric_2}, focusing solely on the first three harmonics. In such figure we also show the sensitivity curves for $\delta \Theta'_{P_b}$ defined in \autoref{eq:varThpPb}. In agreement to what was shown in appendix \autoref{app:beyondPb}, the curves for $\delta \Theta'_{P_b}$ only differ from the ones for $\delta \Theta'$ by order-one factors. The peculiarities related to the linear terms that we mentioned above for $\delta \Psi'$ are also present for $\delta \Theta'$ in the BT model and only the ones with no linear terms resemble those found in the two-step method. The combination of all parameters, primarily influenced by $\delta \Theta'$, for all three high-eccentricity pulsars is depicted in \autoref{fig:more_eccentric_3}, focussing solely on the first three harmonics. 

\subsubsection*{Sensitivity limit (Gaussian priors)} 
We aim to compute $\bar{\mathcal {B}}$ in the case of Rayleigh (for $\varrho_a$) and uniform (for $\Upsilon_a$) priors:
\begin{equation}
P_0 \left( p \right) \deq \Pi_{a=1}^{N_p} P_0 \left( p_a \right)  = \Pi_{a=1}^{N_p} \frac{1}{\pi} e^{-\varrho_a^2} \varrho_a \,.
\end{equation}
In order to compute $\bar{\mathcal {B}}$, we need to compute the integrals over $\varrho_a$, $\Upsilon_a$ and $R^{\mathrm{PN}}_{ai}$. Practically, it is advantageous to first perform the integration over the noise and after that over the variables $X_a \deq \sqrt{2} \varrho_a \cos \Upsilon_a$ and $Y_a \deq \sqrt{2}\varrho_a \sin \Upsilon_a$, leading to two separate Gaussian integrals. For this we decompose $R^{\mathrm{DM}}$ into its $X$ and $Y$ components as
\begin{equation}
    R^{\mathrm{DM}}=\alpha (Q_X X+Q_Y Y)\,,
\end{equation} where
for  the ELL1 case we have 
\begin{equation}
   Q^{\mathrm{ELL1}}_{X(Y)} =   \sin \Psi' A_{X(Y)}^x  + x \cos \Psi' \left[A_{X(Y)}^{\Psi'}+L_{X(Y)}^{\Psi'}\right] - x \frac{\cos(2\Psi')+3}{2} A_{X(Y)}^{\eta} + x \frac{\sin(2 \Psi')}{2} A_{X(Y)}^{\kappa} \,,
\end{equation} 
and for the BT model
{\allowdisplaybreaks
\begin{align}
    Q^{\mathrm{BT}}_{X(Y)}  =&\,\frac{A_{X(Y)}^a}{a}\left[ \eta_b  \sin  E'+\alpha_b\left(\cos E'-e\right)\right]\nonumber\\
    &- A_{X(Y)}^e \left[\alpha_b +\frac{e \eta_b\sin E'  }{1-e^2}+\frac{\sin E'  \left(\alpha_b \sin E'-\eta_b \cos E'\right)}{1-e \cos E'}\right]\nonumber\\
    &+A_{X(Y)}^\omega x\left[ -\sin E' \sin\omega (1-e^2)^{1/2} + \cos \omega \left(\cos E'-e\right)\right]\nonumber\\
    &-\frac{\left[A_{X(Y)}^{\Theta'}+L_{X(Y)}^{\Theta'}\right] \left(\alpha_b  \sin E'-\eta_b  \cos E'\right)}{1-e \cos E'} \,.
\end{align}
}
After integration over noise and $X$ and $Y$ we obtain
\begin{equation}
    \bar{\mathcal {B}} = \exp\left(  \frac{1}{4}\frac{\alpha^4}{\Sigma^4} \right) \,,
\end{equation}
while $\Sigma$ is defined as
\begin{equation}
    \frac{1}{\Sigma^4} \deq \sum_{a=1}^{N_p}  \frac{1}{\Sigma_a^4} \,,
\end{equation}
with
\begin{equation}
    \Sigma_a \deq \frac{1}{\sqrt[4]{2}} \frac{1}{\sqrt{\varrho_{f,a}}} \frac{\epsilon_a}{\sqrt[4]{ \left( \Vec{Q}_{X,a} \cdot \Vec{S}_a \right)^2 + \left( \Vec{Q}_{Y,a} \cdot \Vec{S}_a \right)^2}} \,,
\end{equation}
where the vector $\Vec{S}$ is defined as
\begin{equation}
    \Vec{S} = \sqrt{2} \left(\Vec{Q}_X \cos \Upsilon_f + \Vec{Q}_Y \sin \Upsilon_f \right) \,,
\end{equation}
while in the last equation we have ignored the index $a$, which indicates a particular pulsar. The vectors are indicated by an arrow above the symbols and a dot ($\cdot$) stands for the scalar product. 

Let us compare the sensitivity limits obtained for both priors ($\delta$-function and Gaussian) when considering a single pulsar
\begin{align}
    |\alpha_f|^{\delta} &= \sqrt{\ln\bar{\mathcal{B}}}\, \frac{1}{\varrho_f} \frac{\epsilon}{\sqrt{\Vec{S}^2}} \,,\\
    |\alpha_f|^{G} &= \sqrt[4]{4\ln\bar{\mathcal{B}}}\, \frac{1}{\sqrt[4]{2}} \frac{1}{\sqrt{\varrho_f}} \frac{\epsilon}{\sqrt[4]{ \left( \Vec{Q}_X \cdot \Vec{S} \right)^2 + \left( \Vec{Q}_Y \cdot \Vec{S} \right)^2}} \,.
\end{align}
For $\varrho_f$ large (small) the $\delta$-sensitivity limit outperforms (does not outperform) the Gaussian prior. This is an intuitive result, for example when $\varrho_f$ is large, the signal should be strong, so the $\delta$-prior should produce a low-lying sensitivity line, but a Gaussian prior tends to underestimate this signal by leaning more towards the average. 

\begin{figure}[ht]
     \centering
     \includegraphics[width=13cm]{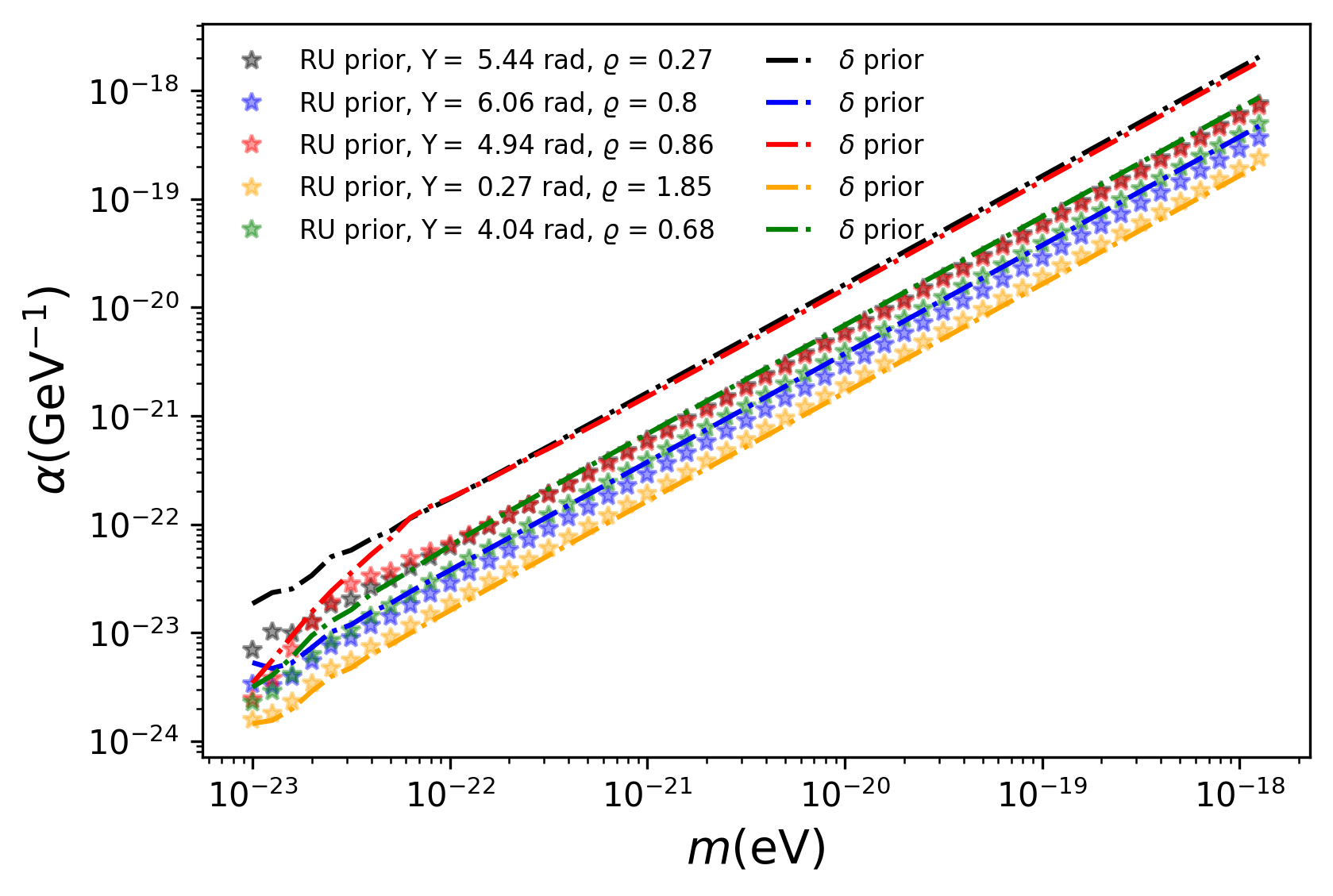}
     \caption{Sensitivity lines obtained from $\delta$- and Gaussian-priors (Rayleigh-Uniform, or ``RU'').}
\label{fig:RUprior_sensitivity_subrang}
\end{figure}

We plot sensitivities obtained from both priors for random realisations of $\varrho_f$ (Rayleigh distribution) and $\Upsilon_f$ (flat distribution). The result is depicted on \autoref{fig:RUprior_sensitivity_subrang}. We can notice that most of the $\delta$-sensitivities are above the sensitivities derived from the Gaussian distribution. This result comes out opposite to the one we arrived at in the main body of the paper. We explain this as follows---for a given value of $\Upsilon_f$ and $m$, we find the value of $\varrho_f^{\mathrm{EQ}}$ so that both sensitivities come out the same
\begin{align}
    \varrho_f^{\mathrm{EQ}} = \sqrt{\frac{\ln \bar{\mathcal{B}}}{2}} F(m,\Upsilon)\,,
\end{align}
with the function $F(m,\Upsilon)$ given as
\begin{equation}
\label{eq:factor}
    F(m,\Upsilon) \equiv \frac{\sqrt{ \left( \Vec{Q}_X \cdot \Vec{S} \right)^2 + \left( \Vec{Q}_Y \cdot \Vec{S} \right)^2}}{\sqrt{\Vec{S}^2}}\,.
\end{equation}
For the method described in the main text, the function $F$ would be different, but the numerical factor $\sqrt{(\ln \bar{\mathcal{B}}) /2}$ is the same. The numerical values may therefore differ slightly, but this is sufficient to change the statistics of the sensitivity lines. For $\bar{\mathcal{B}} = 1000$, \autoref{eq:factor} and 30 values of $\Upsilon$ (on the interval $[0,2\pi)$ with a constant step), we find that $\varrho_f^{\mathrm{EQ}} \ge 1.3$. However, this implies $P(\varrho \ge 1.3) \doteq  0.18$. Let us remind that the median of the distribution is $\varrho = \sqrt{\ln{2}} \doteq 0.833$. Thus, the probability that the sensitivity curve for the $\delta$-prior will be lower than that of the Gaussian priors is less than 18\%. If the factor $F$ was just slightly smaller, it could lead to a probability larger than 50\%. We expect the latter be the case for the method in the main text. 

%-------------------------------------------------------------------------------
%-------------------------------------------------------------------------------
\bibliographystyle{hieeetr}
\bibliography{biblio.bib}

\begin{thebibliography}{10}

\bibitem{Lorimer:2008se}
D.~R. Lorimer, ``{Binary and Millisecond Pulsars},'' {\em Living Rev. Rel.},
  vol.~11, p.~8, 2008, 0811.0762.

\bibitem{Khmelnitsky:2013lxt}
A.~Khmelnitsky and V.~Rubakov, ``{Pulsar timing signal from ultralight scalar
  dark matter},'' {\em JCAP}, vol.~1402, p.~019, 2014, 1309.5888.

\bibitem{Armaleo:2020yml}
J.~M. Armaleo, D.~L\'opez~Nacir, and F.~R. Urban, ``{Pulsar timing array
  constraints on Spin-2 ULDM},'' {\em JCAP}, vol.~09, p.~031, 2020, 2005.03731.

\bibitem{Blas:2016ddr}
D.~Blas, D.~L\'opez~Nacir, and S.~Sibiryakov, ``{Ultralight Dark Matter
  Resonates with Binary Pulsars},'' {\em Phys. Rev. Lett.}, vol.~118, no.~26,
  p.~261102, 2017, 1612.06789.

\bibitem{Blas:2019hxz}
D.~Blas, D.~López~Nacir, and S.~Sibiryakov, ``{Secular effects of ultralight
  dark matter on binary pulsars},'' {\em Phys. Rev. D}, vol.~101, no.~6,
  p.~063016, 2020, 1910.08544.

\bibitem{LopezNacir:2018epg}
D.~L\'{o}pez~Nacir and F.~R. Urban, ``{Vector Fuzzy Dark Matter, Fifth Forces,
  and Binary Pulsars},'' {\em JCAP}, vol.~1810, no.~10, p.~044, 2018,
  1807.10491.

\bibitem{Armaleo:2019gil}
J.~M. Armaleo, D.~L\'opez~Nacir, and F.~R. Urban, ``{Binary Pulsars as probes
  for Spin-2 Ultralight Dark Matter},'' {\em JCAP}, vol.~2001, no.~01, p.~053,
  2020, 1909.13814.

\bibitem{Manchester:2004bp}
R.~N. Manchester, G.~B. Hobbs, A.~Teoh, and M.~Hobbs, ``{The Australia
  Telescope National Facility pulsar catalogue},'' {\em Astron. J.}, vol.~129,
  p.~1993, 2005, astro-ph/0412641.
\newblock Available at \url{https://www.atnf.csiro.au/research/pulsar/psrcat/}.

\bibitem{Lorimer:2004}
D.~R. {Lorimer} and M.~{Kramer}, {\em Handbook of Pulsar Astronomy}.
\newblock Vol.~4.~Cambridge, UK: Cambridge University Press, 2004, Dec. 2004.

\bibitem{Teukolsky1976}
R.~Blandford and S.~A. Teukolsky, ``{Arrival-time analysis for a pulsar in a
  binary system},'' {\em Astrophysical Journal}, vol.~205, pp.~580--591, 1976.

\bibitem{Lange:2001rn}
C.~Lange, F.~Camilo, N.~Wex, M.~Kramer, D.~C. Backer, A.~G. Lyne, and
  O.~Doroshenko, ``{Precision timing measurements of psr j1012+5307},'' {\em
  Mon. Not. Roy. Astron. Soc.}, vol.~326, p.~274, 2001, astro-ph/0102309.

\bibitem{Hobbs:2006cd}
G.~Hobbs, R.~Edwards, and R.~Manchester, ``{Tempo2, a new pulsar timing
  package. 1. overview},'' {\em Mon. Not. Roy. Astron. Soc.}, vol.~369,
  pp.~655--672, 2006, astro-ph/0603381.

\bibitem{Foster:2017hbq}
J.~W. Foster, N.~L. Rodd, and B.~R. Safdi, ``{Revealing the Dark Matter Halo
  with Axion Direct Detection},'' {\em Phys. Rev. D}, vol.~97, no.~12,
  p.~123006, 2018, 1711.10489.

\bibitem{Danby:1970}
J.~Danby, {\em {Fundamentals of Celestial Mechanics}}.
\newblock MacMillan, 1970.

\bibitem{Moore:2014eua}
C.~J. Moore, S.~R. Taylor, and J.~R. Gair, ``{Estimating the sensitivity of
  pulsar timing arrays},'' {\em Class. Quant. Grav.}, vol.~32, no.~5,
  p.~055004, 2015, 1406.5199.

\bibitem{Luo:2020ksx}
J.~Luo {\em et~al.}, ``{PINT: A Modern Software Package for Pulsar Timing},''
  {\em Astrophys. J.}, vol.~911, no.~1, p.~45, 2021, 2012.00074.

\bibitem{nanograv2023}
G.~A. et~al, ``{The NANOGrav 15 yr Data Set: Observations and Timing of 68
  Millisecond Pulsars},'' {\em ApJL}, vol.~951, p.~78, 2023, 2306.16217.
\newblock Available at \url{https://zenodo.org/records/8423265}.

\bibitem{Bertotti:2003rm}
B.~Bertotti, L.~Iess, and P.~Tortora, ``{A test of general relativity using
  radio links with the Cassini spacecraft},'' {\em Nature}, vol.~425,
  pp.~374--376, 2003.

\bibitem{Armstrong:2003ay}
J.~W. Armstrong, L.~Iess, P.~Tortora, and B.~Bertotti, ``{Stochastic
  gravitational wave background: Upper limits in the 10**-6-Hz 10**-3-Hz
  band},'' {\em Astrophys. J.}, vol.~599, pp.~806--813, 2003.

\bibitem{Porayko:2018sfa}
N.~K. Porayko {\em et~al.}, ``{Parkes Pulsar Timing Array constraints on
  ultralight scalar-field dark matter},'' {\em Phys. Rev.}, vol.~D98, no.~10,
  p.~102002, 2018, 1810.03227.

\bibitem{EuropeanPulsarTimingArray:2023egv}
C.~Smarra {\em et~al.}, ``{Second Data Release from the European Pulsar Timing
  Array: Challenging the Ultralight Dark Matter Paradigm},'' {\em Phys. Rev.
  Lett.}, vol.~131, no.~17, p.~171001, 2023, 2306.16228.

\bibitem{Liu:2011cka}
K.~Liu, J.~P.~W. Verbiest, M.~Kramer, B.~W. Stappers, W.~van Straten, and J.~M.
  Cordes, ``{Prospects for High-Precision Pulsar Timing},'' {\em Mon. Not. Roy.
  Astron. Soc.}, vol.~417, p.~2916, 2011, 1107.3086.

\bibitem{Shevchuk:2023ccb}
T.~Shevchuk, E.~D. Kovetz, and A.~Zitrin, ``{New Bounds on Fuzzy Dark Matter
  from Galaxy-Galaxy Strong-Lensing Observations},'' 8 2023, 2308.14640.

\bibitem{Aghaie:2023lan}
M.~Aghaie, G.~Armando, A.~Dondarini, and P.~Panci, ``{Bounds on Ultralight Dark
  Matter from NANOGrav},'' 8 2023, 2308.04590.

\bibitem{Stott:2020gjj}
M.~J. Stott, ``{Ultralight Bosonic Field Mass Bounds from Astrophysical Black
  Hole Spin},'' {\em preprint (arxiv: 2009.07206)}, Sep 2020, 2009.07206.

\bibitem{Armaleo:2020efr}
J.~M. Armaleo, D.~L\'opez~Nacir, and F.~R. Urban, ``{Searching for spin-2 ULDM
  with gravitational waves interferometers},'' {\em JCAP}, vol.~04, p.~053,
  2021, 2012.13997.

\bibitem{Turner:1979yn}
M.~S. Turner, ``{Influence of a weak gravitational wave on a bound system of
  two point masses},'' {\em Astrophys. J.}, vol.~233, pp.~685--693, 1979.

\bibitem{watson1995treatise}
G.~Watson, {\em A Treatise on the Theory of Bessel Functions}.
\newblock Cambridge Mathematical Library, Cambridge University Press, 1995.

\bibitem{1970SvA....13..562S}
I.~S. {Shklovskii}, ``{Possible Causes of the Secular Increase in Pulsar
  Periods.},'' {\em Soviet Astronomy}, vol.~13, p.~562, Feb. 1970.

\bibitem{1991ApJ...366..501D}
T.~{Damour} and J.~H. {Taylor}, ``{On the orbital period change of the binary
  pulsar PSR 1913 + 16},'' {\em Astrophysical Journal}, vol.~366, pp.~501--511,
  Jan. 1991.

\end{thebibliography}
\end{document}